\documentclass[
 reprint,
superscriptaddress,
 amsmath,amssymb,
 aps,
prl,
]{revtex4-2}

\usepackage{xcolor}
\usepackage{dsfont}
\usepackage{mathtools}
\usepackage{braket}
\renewcommand{\ket}[1]{\mathop{\lvert#1\rangle}}
\renewcommand{\bra}[1]{\mathop{\langle#1\rvert}}

\DeclarePairedDelimiterX{\norm}[1]{\lVert}{\rVert}{#1}
\usepackage{graphicx}
\usepackage{dcolumn}
\usepackage{bm}
\usepackage[colorlinks = true,
            linkcolor = blue,
            urlcolor  = blue,
            citecolor = blue,
            anchorcolor = blue]{hyperref}
\DeclareMathOperator{\Tr}{tr}
\usepackage{float}

\begin{document}

\preprint{APS/123-QED}

\title{Nonstabilizerness and Error Resilience in Noisy Quantum Circuits}

\author{Fabian Ballar Trigueros}
\email{fabian.ballar@uni-a.de}
\affiliation{%
Theoretical Physics III, Center for Electronic Correlations and Magnetism,
Institute of Physics, University of Augsburg, 86135 Augsburg, Germany
}%

\author{Jos\'e Antonio Mar\'in Guzm\'an}
\email{marin@umd.edu}
\affiliation{%
Joint Center for Quantum Information and Computer Science, NIST and University of Maryland,
College Park, MD 20742, United States of America
}%

\date{\today}

\begin{abstract}
\noindent We investigate how noise impacts nonstabilizerness—a key resource for quantum advantage—in many-body qubit systems. While noise typically degrades quantum resources, we show that amplitude damping, a nonunital channel, can generate or enhance magic, whereas depolarizing noise provably cannot. In an encoding-decoding protocol, we find that, unlike in the coherent-noise case, a sharp decoding fidelity transition is not accompanied by a transition in nonstabilizerness. Although amplitude damping locally injects magic, this resource is washed out at the collective level after encoding, decoding, and postselection. Our results reveal that realistic incoherent noise can suppress many-body magic criticality even while generating it microscopically.
\end{abstract}

\maketitle


Understanding how quantum properties degrade---or, in some cases, persist---under realistic noise is a central challenge in quantum information. Among the various signatures of quantumness, nonstabilizerness---or magic---has emerged as a key resource: it quantifies how far a state deviates from the set of stabilizer states, which can be efficiently simulated classically~\cite{Gottesman98, Gottesman24}. Nonstabilizerness is not only central to universal quantum computation~\cite{Veitch14, Howard_2017}, but also informs the study of many-body quantum systems~\cite{Liu22, White21,Tarabunga2024critical,magni2025}, noise thresholds, and the capabilities of quantum devices in the noisy intermediate-scale quantum era~\cite{Preskill18,Nelson24,wei2024noiserobustnessthresholdmanybody} and beyond.

Several approaches have been proposed to quantify magic in qubit systems, which are central to most experimental platforms and subject to physically motivated noise \cite{Chirolli08, Blais16, Ghosh12}. Stabilizer Rényi entropies (SREs)~\cite{Leone22, Haug_2023} offer a proxy that can be computed efficiently even in large-scale simulations. However, SREs are not monotones for mixed states ~\cite{Leone_2024, Haug_2023, Tarabunga2024critical}. A more rigorous alternative is the robustness of magic (ROM)~\cite{Howard_2017, Heinrich_2019}, a faithful resource monotone that is nonincreasing under stabilizer-preserving operations, though it becomes demanding to compute for large systems due to the superexponential number of stabilizer states~\cite{10.1063/1.2393152, Hamaguchi_2024, PhysRevApplied.23.014069}.

Here, we investigate how physically relevant incoherent noise affects nonstabilizerness in many-body qubit systems. While noise is often expected to suppress quantum resources, we show that this expectation is not universal: amplitude damping, a paradigmatic nonunital channel, can generate or enhance magic, whereas depolarizing noise universally suppresses magic and cannot generate it. This immediately reveals that realistic incoherent noise channels can differ qualitatively in their effect on nonstabilizerness, even at the single-state level.

As a concrete many-body setting, we study an encoding-decoding protocol subjected to amplitude damping noise. Previous works established that such circuits exhibit an error-resilience transition in the decoding fidelity~\cite{PhysRevLett.132.140401}, while coherent noise can induce a transition in the magic content of the decoded state~\cite{Niroula24,sierant2026magic}. We show that under amplitude damping, the decoding fidelity still undergoes a sharp transition, but no corresponding transition occurs in the robustness of magic. Extending this analysis to competing coherent and incoherent errors, we find that the error-resilient phase survives only within a bounded region of parameter space, while the robustness of magic displays a clear crossover in regimes where coherent noise alone would generate extensive nonstabilizerness. These results demonstrate that local magic generation does not necessarily imply collective many-body magic criticality: amplitude damping, although itself capable of generating magic, competes with coherent noise as an effective magic-depleting process that suppresses the buildup of extensive nonstabilizerness.

\begin{figure*}
    \centering
    \includegraphics[width=\linewidth]{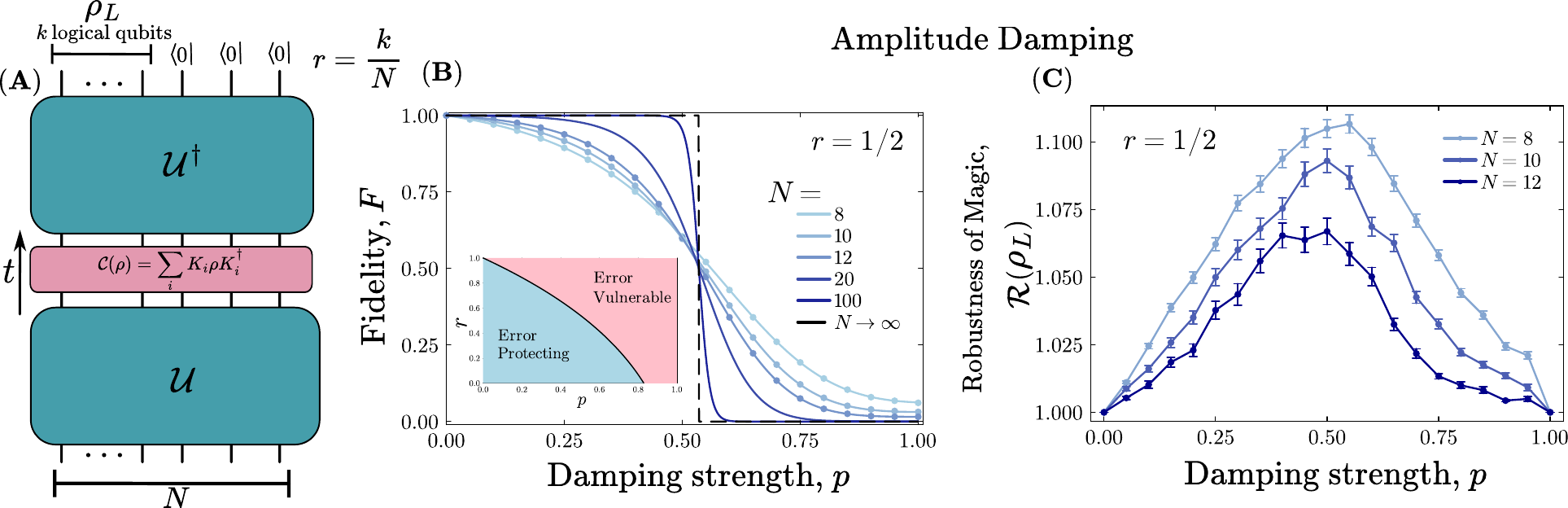}
    \caption{(\textbf{A}) Schematic of the encoding-decoding protocol. A quantum state consisting of $k$ logical qubits is embedded into a larger system of $N$ physical qubits using a unitary operation. After the system undergoes noise, a decoding unitary is applied to attempt recovery of the logical information. The protocol is post-selected on outcomes where the ancillary qubits are found in the zero state. (\textbf{B}) Fidelity of the recovered state as a function of the damping parameter $p$, showing a transition from error-protecting to error-vulnerable behavior. Curves show analytical predictions, while data points correspond to numerical simulations. The inset highlights the boundary between these two phases. (\textbf{C}) Average robustness of magic under amplitude damping within the encoding-decoding framework. ROM values are averaged over 1000, 500, and 300 Clifford unitary realizations for increasing $N$.}
    \label{fig:One}
\end{figure*}

\textit{Quantifying nonstabilizerness.—}Quantum circuits composed of Clifford gates, generated by $S = \sqrt{Z}$, $H = (X+Z)/\sqrt{2}$, and CNOT, can be efficiently simulated classically. Any state obtained by applying such a circuit to  $|0\rangle^{\otimes N}$ is a stabilizer state. To achieve universal quantum computation, non-Clifford resources are required \cite{Gottesman98}. The extent to which a quantum state deviates from a stabilizer state constitutes a computational resource, known as nonstabilizerness, or magic.

Any quantum state $\rho$ can be expanded as $\rho = \sum_i c_i |\psi_i\rangle\langle\psi_i|$, where ${|\psi_i\rangle}$ are stabilizer states forming an overcomplete basis of the Hermitian operator space. States within the stabilizer polytope satisfy $c_i \geq 0$, $\sum_i c_i = 1$, and are nonmagic. Magic states, by contrast, include negative components in any such decomposition. The degree of this negativity, captured by the minimal total weight $\sum_i|c_i|$, defines the state's robustness of magic.

Let $\mathcal{S}_N$ denote the set of $N$-qubit stabilizer states. ROM \cite{Howard_2017} is defined as
\begin{equation}
\mathcal{R}(\rho) = \min\left\{ \sum_i |x_i| \;\middle|\; \rho = \sum_{i} x_i |\psi_i\rangle\langle\psi_i|,\; |\psi_i\rangle \in \mathcal{S}_N \right\},
\end{equation}
where the minimization runs over all decompositions into stabilizer states $|\psi_i\rangle$. For stabilizer states, $\mathcal{R}(\rho) = 1$; for magic states, $\mathcal{R}(\rho) > 1$, quantifying their distance from the convex hull of stabilizer states, known as the stabilizer polytope \cite{Heinrich_2019}.

ROM satisfies several key properties \cite{Heinrich_2019}:  
(i) \textit{Faithfulness}: $\mathcal{R}(\rho) = 1$ if and only if $\rho$ is a stabilizer state;  
(ii) \textit{Submultiplicativity}: $\mathcal{R}(\rho_1 \otimes \rho_2) \leq \mathcal{R}(\rho_1) \mathcal{R}(\rho_2)$;  
(iii) \textit{Monotonicity}: ROM does not increase under stabilizer-preserving quantum channels, i.e., $\mathcal{R}(\mathcal{E}(\rho)) \leq \mathcal{R}(\rho)$;  
(iv) \textit{Convexity}: For any ensemble $\{p_k, \rho_k\}$, $\mathcal{R}(\sum_k p_k \rho_k) \leq \sum_k p_k \mathcal{R}(\rho_k)$.

$\mathcal{S}_N$ grows superexponentially with system size as $|\mathcal{S}_N| = 2^N \prod_{i=1}^N (2^i+1)$~\cite{10.1063/1.2393152}, making ROM computation challenging. Brute-force methods are limited to $N \leq 5$, while more advanced linear-programming approaches, such as column generation~\cite{Hamaguchi_2024}, enable exact evaluations up to $N = 8$ qubits. For larger systems, computing ROM becomes infeasible. Despite its considerable computational cost, the ROM constitutes an optimal measure for quantifying nonstabilizerness, particularly for mixed states \cite{leone2026}.

\textit{Noise channels.—}Noise in quantum devices is intricate, and its accurate modeling presents a significant challenge. A quantum channel can be expressed as a completely positive trace-preserving (CPTP) operation, defined in terms of Kraus operators as $\mathcal{C}(\rho) = \sum_{i} K_i \rho K_i^{\dagger}$, subject to the constraint $\sum_{i} K_i^{\dagger} K_i = \mathds{1}$, which ensures trace preservation \cite{Nielsen10}. 

We study noise through two paradigmatic models: depolarizing noise and amplitude damping. The depolarizing channel is an example of unital noise, a channel that maps the identity matrix to itself. We show that an instance of nonunital noise, amplitude damping, has a nontrivial effect on quantum states' nonstabilizerness.

The depolarizing channel, $ \mathcal{D}_p(\rho) = \frac{p}{2^N}\mathds{1} + (1 - p)\rho $, describes a mixture of the input state and the maximally mixed state; $ p $ tunes the noise strength. Depolarizing noise, being a convex mixture of Pauli operations, cannot generate magic.  The convexity of ROM ensures that $\mathcal{R}(\mathcal{D}_p(\rho)) \leq (1-p)\, \mathcal{R}(\rho) + p$, implying monotonic decay for magic states and constancy at one for stabilizer states.

\begin{figure*}
    \centering
    \includegraphics[width=\linewidth]{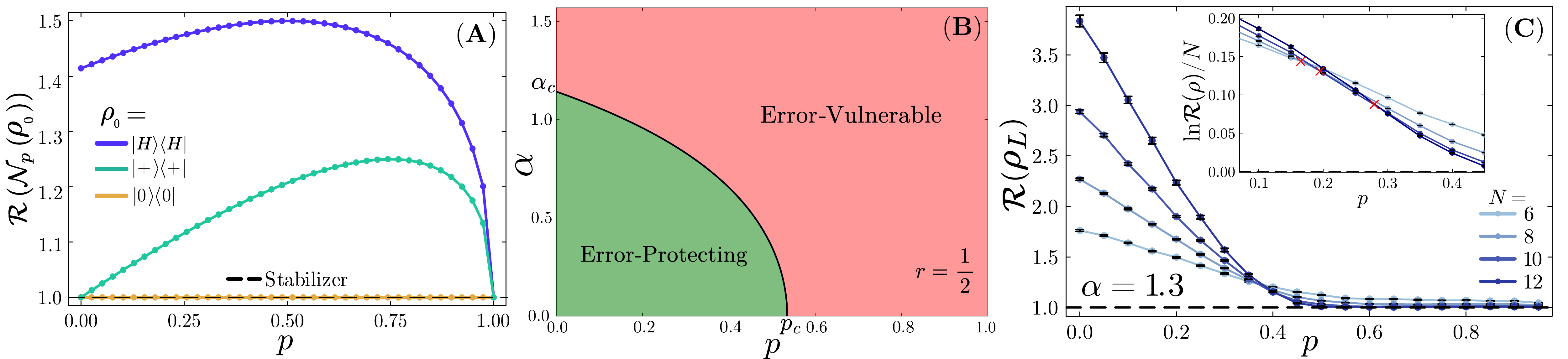}
    \caption{(\textbf{A}) ROM for various initial states under amplitude damping noise versus damping strength $p$. (\textbf{B}) Phase diagram at fixed code rate $r=1/2$ of the competing coherent error and amplitude damping. (\textbf{C}) ROM as a function of the amplitude damping parameter $p$ at a fixed coherent error strength $\alpha = 1.3$. The inset shows the normalized log ROM near the cross point; the red X marks the crossings' shifting due to finite-size effects.}
    \label{fig:Two}
\end{figure*}

To study a physically relevant incoherent channel capable of generating magic, we consider the amplitude-damping (AD) channel, which describes irreversible relaxation toward the ground state and provides a paradigmatic example of nonunital noise~\cite{Nielsen10,Khatri20}. Its single-qubit Kraus operators are

\begin{equation} 
\begin{aligned} K_0 &= \begin{pmatrix} 1 & 0 \\ 0 & \sqrt{1{-}p} \end{pmatrix} \quad\text{ and }\quad 
K_1 = \begin{pmatrix} 0 & \sqrt{p} \\ 0 & 0 \end{pmatrix} .\end{aligned} \label{Eq:Kraus} \end{equation}
The channel acts as $\mathcal{N}_p(\rho)=\sum_{i=0}^{1} K_i \rho K_i^\dagger$; $p$ controls the damping strength. Writing a qubit state as $\rho=\tfrac12(\mathds{1}+r_x X+r_y Y+r_z Z)$, the Bloch vector transforms as $\mathbf r \mapsto (r_x\sqrt{1-p},, r_y\sqrt{1-p},, p+r_z(1-p))$, showing that the Bloch sphere contracts toward the north pole $|0\rangle$. This displacement of the sphere's center distinguishes amplitude damping from unital channels such as depolarizing noise and underlies its ability to generate nonstabilizerness. A generalized finite-temperature extension of this channel is discussed in the Supplemental Material.

\textit{Nonstabilizerness and noise.—}Incoherent noise is often expected to degrade quantum resources by driving quantum states toward classical mixtures. This is indeed the case for depolarizing noise, which cannot generate magic and instead monotonically suppresses it for any input state. A natural question is therefore whether this depletion of nonstabilizerness is a universal property of noisy dynamics, or whether physically relevant channels can instead enhance a state's quantumness. We find that amplitude damping, despite being incoherent and dissipative, can both generate and enhance magic, showing that nonunital noise can inject nonstabilizerness into quantum systems.

This effect already appears at the single-qubit level. Figure~\ref{fig:Two}(\textbf{A}) shows the robustness of magic of $|+\rangle$, $|H\rangle=(|0\rangle+e^{i\pi/4}|1\rangle)/\sqrt{2}$, and $|0\rangle$ under amplitude damping. The channel generates magic from stabilizer inputs such as $|+\rangle$, while $|0\rangle$ remains unchanged as the fixed point of the dynamics. For nonstabilizer inputs such as $|H\rangle$, the robustness of magic is initially enhanced before eventually decreasing at stronger damping. Geometrically, amplitude-damping trajectories flow toward $|0\rangle$, and, for suitable initial conditions, traverse outside the stabilizer polytope before relaxing to the fixed point.

To illustrate that nonunital noise can also be operationally useful, we note that amplitude damping combined with postselection on no-click trajectories can probabilistically generate pure magic states using otherwise stabilizer ingredients. In particular, starting from a Bell pair of a system qubit and an ancilla, conditioning on no decay and measuring the system qubit yields the ancilla state $|\Psi_p\rangle=(|0\rangle+\sqrt{1-p}|1\rangle)/\sqrt{2-p}$, which is nonstabilizer for all $p\neq 0,1$. A detailed derivation and geometric interpretation are given in the End Matter.

The key question is whether this local noise-induced magic survives as a collective resource in many-body protocols.

\textit{Encoding-decoding circuits.—}The error-resilience transition, introduced in Refs.~\cite{PhysRevLett.132.140401}, provides a probe of how well quantum codes withstand noise. It manifests as a sharp change in the fidelity of a decoding protocol: below a critical threshold, logical information can be successfully recovered, while above it decoding fails.

In Fig.\,\ref{fig:One}(\textbf{A}), we illustrate the noisy encoding-decoding circuit. We initialize an $N$-qubit state $\rho_0 = |0\rangle\langle 0|_L \otimes |0\rangle\langle 0|_A$, where the first register corresponds to a $k$-qubit logical state and the second consists of ancilla qubits used for syndrome measurement during error correction, with code rate $r = k/N$. To implement a global Clifford unitary, we use a depth-$4N$ circuit of alternating layers: odd layers apply uniformly random single-qubit Clifford gates, and even layers apply $XX(\pi/4)$ rotations on $N/2$ randomly chosen qubit pairs. This structure follows Ref.~\cite{Niroula24}, with the addition of nonunital noise and postselection. We denote by $U$ the noiseless unitary circuit.

After the unitary evolution, the state undergoes local noise described by $\mathcal{C} = \bigotimes_{i=1}^N \mathcal{C}_i$, where each $\mathcal{C}_i$ acts independently on qubit $i$. Explicitly, $\mathcal{C}(\rho) = \sum_{\vec{j}} K_{\vec{j}} \rho K_{\vec{j}}^\dagger$ with $K_{\vec{j}} = \bigotimes_{i} K_{j_i}^{(i)}$, where $\vec{j} = (j_1, j_2, \dots, j_N)$ and $K_{j_i}^{(i)}$ is a Kraus operator of $\mathcal{C}_i$. Decoding proceeds by reversing the unitary, $\rho = U^\dagger \mathcal{C}(U \rho_0 U^\dagger) U$, followed by measuring all ancillas in the computational basis and post-selecting on the outcome $|0\rangle^{\otimes (N-k)}$ \footnote{Projective ancilla measurements are also viable, but their outcome distribution may itself undergo a transition. To avoid additional numerical noise, we post-select on a fixed outcome.}. The logical state is then given by $\rho_L = \langle 0_A | \rho | 0_A \rangle / \mathrm{Tr}(\langle 0_A | \rho | 0_A \rangle)$.

In Fig.\,\ref{fig:One}(\textbf{B–C}), we show the decoding fidelity and the robustness of magic as functions of the amplitude-damping parameter~$p$. To calculate the fidelity, we use an analytical approach based on the replica trick. This follows the method developed in Ref.~\cite{PhysRevLett.132.140401}. We find an encoding-decoding phase transition at a critical  value $p_c =   2^{(3+r)/2}-2^{r+1} $. The fidelity under amplitude-damping noise is
\begin{equation}
    F_{\textrm{AD}} = \frac{(2^N - 1) \left[ 1 + (1 + \sqrt{1-p} - p/2)^N \right]}{(2^N - 2^k)\left[ 1 + \sqrt{1-p} - p/2 \right]^N + 2^{N+k} - 1}.
    \label{Eq:FAD}
\end{equation}

We provide further details of this calculation in the SM \footnote{We note that Eq.~\eqref{Eq:FAD} results from an annealed average; numerics (see SM) confirm that the quenched fidelity closely follows this expression for the sizes studied, and the difference decreases with $N$.}. 
 Since we begin in a stabilizer state, this transition implies that below the critical threshold, the system generates no magic, even in the thermodynamic limit. The natural question, then, is whether the system develops an increasing amount of magic for $p > p_c$, potentially signaling a corresponding phase transition in nonstabilizerness.

The magic of mixed states is notoriously difficult to study, lacking both tractable analytical forms and straightforward numerical treatments. We employ the column-generation algorithm developed in \cite{Hamaguchi2024ROM} to compute the ROM averaged over trajectories. As shown in Fig.\,\ref{fig:One}(\textbf{C}), the robustness of magic decreases with increasing system size. In contrast, if a genuine transition in nonstabilizerness were present, one would expect the ROM to grow with system size, reflecting the buildup of extensive many-body magic. This increase would lead to a crossing point in the different-size curves; the presence of size-independent behavior would signal criticality. Thus, even in the presence of a magic-generating channel, the system shows no evidence of a phase transition in nonstabilizerness.

We analyze the protocol through its effective logical channel. For each encoder $U$, the noise–decode–postselect map yields a normalized logical state $\rho_U$. One may either average the robustness directly, $\mathbb{E}_U[\mathcal{R}(\rho_U)]$, or evaluate the robustness of the mean state, $\mathcal{R}(\bar{\rho})$ with $\bar{\rho}=\mathbb{E}_U[\rho_U]$. By convexity of ROM, $\mathcal{R}(\bar{\rho})\le \mathbb{E}_U[\mathcal{R}(\rho_U)]$. Averaging the channel over encoders projects it onto the commutant of logical unitaries\cite{Nielsen_2002, Emerson_2005, horodecki1998}, giving a depolarizing map on the logical space of dimension $d_L=2^k$,
\begin{equation}
\bar{\Lambda}^{\mathrm{cond}}(\rho)=\tau\,\rho+(1-\tau)\,\tfrac{\mathds{1}}{d_L},\qquad 
\tau=\frac{d_L F_{\mathrm{AD}}-1}{d_L-1},
\label{Eq:Taumt}
\end{equation}
where $F_{\mathrm{AD}}$ is the decoding fidelity in Eq.\,\eqref{Eq:FAD}. Since the depolarized output lies in the stabilizer polytope for stabilizer inputs, this fixes $\mathcal{R}(\bar{\rho})=1$ for all $p$. Moreover, ROM is Lipschitz-continuous in trace norm~\cite{Schluck_2023}; thus, if the post-selected logical states concentrate around the mean, $\|\rho_U - \bar{\rho}\|_1 \xrightarrow{N \to \infty} 0$,then
\begin{equation}
0\le \mathbb{E}_U[\mathcal{R}(\rho_U)]-\mathcal{R}(\bar{\rho}) \;\le\; L(d_L)\,\mathbb{E}_U\!\big[\|\rho_U-\bar{\rho}\|_1\big]\xrightarrow{N\to\infty}0,
\end{equation}
forcing $\mathbb{E}_U[\mathcal{R}(\rho_U)]\to 1$. Intuitively, amplitude damping plus postselection can inject magic into individual trajectories, but random encoding symmetrizes these outputs so that, after decoding and conditioning, the ensemble contracts onto the depolarizing ray set by $F_{\mathrm{AD}}$; the fluctuations that carry magic are washed out. The End Matter and SM provide analytic bounds and numerics supporting this concentration mechanism.

This mechanism shows that the absence of a magic transition is a structural consequence of incoherent noise combined with postselection. For amplitude damping the ensemble of post-selected states concentrates onto a depolarizing channel, asymptotically washing out any resource transition. In contrast, for the coherent-noise case, we show in the End Matter that concentration fails: fluctuations (i.e., the spread of $\rho_U$ around $\bar{\rho}$) persist precisely where magic displays a transition, consistent with the findings of Ref.~\cite{Niroula24, sierant2026magic}. As a result, the decoding transition in fidelity need not coincide with a transition in nonstabilizerness.

Amplitude damping can generate magic in individual trajectories, but after encoding, decoding, and postselection, the ensemble asymptotically concentrates onto a depolarizing channel that carries no resource. The absence of a transition in ROM therefore, does not mean that single realizations are classically simulable, but that fluctuations average out in the thermodynamic limit. Whether such fluctuations survive or collapse provides a natural way to distinguish noise-driven resource phase transitions in quantum circuits.

\textit{Encoding-decoding with competing noise.—}A natural question is whether the decoupling between transitions in pure amplitude damping is special to that channel. We therefore consider a mixed local error model combining incoherent and coherent errors. On each qubit we first apply amplitude damping and subsequently a coherent $Z$ rotation, such that the local channel reads $\mathcal{C}_{p,\alpha}(\rho)=e^{-i\alpha Z/2}\,\mathcal{N}_p(\rho)\,e^{i\alpha Z/2}$, where $p$ denotes the damping strength and $\alpha$ the coherent rotation angle. Equivalently, this corresponds to modified Kraus operators $\widetilde K_i=e^{-i\alpha Z/2}K_i$, with $K_i$ the amplitude-damping Kraus operators introduced above. The encoding, decoding, and postselection protocol is otherwise unchanged.

The decoding fidelity can again be obtained analytically using the same replica and Weingarten framework as in the pure amplitude-damping case. The only modification is that the corresponding single-site transfer factor becomes $B(p,\alpha)=2-p+2\sqrt{1-p}\cos\alpha$, yielding the annealed decoding fidelity
\begin{equation}
F_{p,\alpha}
=
\frac{(2^N-1)\left(2^N+B(p,\alpha)^N\right)}
{2^N\left(2^{k+N}-1\right)+\left(2^N-2^k\right)B(p,\alpha)^N}.
\label{Eq:Fmix}
\end{equation}
As in the amplitude-damping case, numerics show that the quenched fidelity closely follows Eq.~\eqref{Eq:Fmix} for the accessible system sizes. The corresponding critical curve follows from the competition between exponential contributions in Eq.~\eqref{Eq:Fmix}, namely $2-p+2\sqrt{1-p}\cos\alpha = 2^{1+r}$, and defines the phase boundary shown in Fig.~\ref{fig:Two}(\textbf{B}). The error-resilient phase survives only within a bounded region of the $(p,\alpha)$ plane: once either coherent or incoherent noise exceeds threshold, increasing the other no longer restores successful decoding.

We then examine the ROM along cuts through this mixed-noise regime. Fixing $\alpha$ deep in the coherent-noise magic-generating regime, Fig.~\ref{fig:Two}(\textbf{C}) shows that increasing amplitude damping suppresses the growth of nonstabilizerness and instead produces a clear size-dependent crossover toward trivial robustness values. Nevertheless, we cannot establish whether this crossover is asymptotically sharp. Importantly, this behavior occurs in a regime where the decoding fidelity is already in the error-vulnerable phase, while coherent noise alone would generate extensive nonstabilizerness \cite{sierant2026magic}. This provides further evidence that signatures of many-body magic can emerge independently of error-resilience transitions, deepening the decoupling between the two phenomena.

At the same time, the presence of coherent errors shows that magic transitions are not generically destroyed by incoherent noise, but rather reshaped by the competition between noise mechanisms. Thus, although amplitude damping can generate magic locally, in competition with coherent noise it acts effectively as a magic-depleting process: the local resource injected by the channel does not survive collective averaging, and extensive nonstabilizerness is progressively suppressed.

\textit{Conclusion and outlook.—}
In this work, we investigated how physically relevant noise affects quantum nonstabilizerness in many-body qubit systems. Using the robustness of magic, we showed that nonunital channels such as amplitude damping can generate or enhance magic, in contrast to depolarizing noise, which strictly depletes it. This demonstrates that realistic incoherent noise can differ qualitatively in its effect on quantum-computational resources.

We applied these insights to encoding-decoding transitions in noisy quantum circuits. While this setting exhibits a sharp transition in decoding fidelity, we found no corresponding transition in nonstabilizerness under amplitude damping, even though the channel locally generates magic. Our analysis shows that this behavior arises from a concentration mechanism: after encoding, decoding, and postselection, the ensemble of logical states collapses onto an effective depolarizing channel. 

These results establish a key conceptual point: local resource generation does not necessarily imply collective many-body resource criticality. In contrast to coherent errors, where fidelity and magic transitions coincide~\cite{sierant2026magic}, realistic incoherent noise can suppress the buildup of extensive nonstabilizerness even while generating it microscopically.

Our findings open several directions, including the possibility of harnessing nonunital noise as a resource, the development of scalable and faithful diagnostics of mixed-state magic, and a deeper understanding of how noise shapes quantum resources in many-body systems. More broadly, our results show that the relationship between noise, error-resilience, and quantum-computational resources is not universal, but depends sensitively on the structure of the noise, highlighting the need for refined frameworks to classify noise-driven resource transitions in quantum circuits.

\section*{Data availability}
The data contained in all figures of this article is available on Zenodo \cite{Ballar_Marin}.

\vspace{0.5cm}
\textit{Acknowledgments—}We thank L. Brunner, M. Gullans, X. Turkeshi, C.D. White, and N. Yunger Halpern for insightful discussions. The code for our implementation of SRE and ROM can be found on GitHub \cite{Ballar2025}. We also thank H. Hamaguchi for his implementation of the column-generating algorithm for the ROM \cite{Hamaguchi2024ROM}. FBT has received funding from the European Research Council (ERC) under the European Union’s Horizon 2020 research and innovation programme (grant agreement No. 853443). FBT gratefully acknowledges the resources on the LiCCA HPC cluster of the University of Augsburg, co-funded by the Deutsche Forschungsgemeinschaft (DFG, German Research Foundation)-Project-ID 499211671. We thank the University of Costa Rica, where this collaboration began.

\clearpage
\section{End Matter}

\noindent
\textit{Stabilizer Renyi Entropies.—}An alternative measure of magic is the stabilizer R\'enyi entropy \cite{Leone22}. The Pauli group $\mathcal{P}_N$ consists of tensor products of $\mathds{1}, X, Y,$ and $Z$ on $N$ qubits, with an overall phase of $\pm1$ or $\pm i$. For an $N$-qubit state $\rho$, the SRE is 

\begin{equation}
    \widetilde{\mathcal{M}}_\alpha(\rho) = \frac{1}{1 - \alpha} \left[ \ln\left(\sum_{P \in \mathcal{P}_n} \frac{|\operatorname{tr}(\rho P)|^{2\alpha}}{2^N} \right) -\ln \mathrm{tr}(\rho^2) \right].
    \label{eq:SRE}
\end{equation}

\noindent
SREs possess some desirable properties for quantifying pure-state nonstabilizerness: they are faithful and remain invariant under stabilizer operations.

This calculation is significantly more efficient than the superexponential complexity of the ROM. New techniques are capable of computing the SRE for systems of up to $24$ qubits \cite{Turkeshi_2025, Sierant26Comp}. Furthermore, Markov chain methods, neural quantum state, and tensor network protocols have been developed to improve computational efficiency \cite{PRXQuantum.4.040317,PhysRevLett.133.010601,sinibaldi2025,Turkeshi_2025,PhysRevB.111.054301,PhysRevA.111.012440,PhysRevLett.133.010602,pavi2024,PhysRevLett.131.180401}.

For mixed states, this quantity can reliably certify a stabilizer state only when the state admits a purification that is itself a stabilizer state \cite{Tarabunga2024critical,ding2025evaluatingmanybodystabilizerrenyi,tarabunga2025efficientmutualmagicmagic}; SREs do not represent faithful measures of magic for general mixed states. An extension using convex roof constructions has been proposed, but it requires an optimization procedure impractical for large systems \cite{Leone_2024}.\\

\noindent
\textit{Effective Channel and Concentration.—}
As discussed in the main text, the apparent mismatch that amplitude damping can generate magic locally (Fig.\,\ref{fig:Two}\textbf{A}) but not in the encoding–decoding setup (Fig.\,\ref{fig:One}\textbf{C}) is resolved by analyzing the effective logical channel. 
For each encoder $U$, the noisy, decoded, and post-selected map can be written as a trace-decreasing channel 
$\Lambda_U(\rho_0) = \sum_{\vec j} M_{\vec j}(U)\rho_L M_{\vec j}^\dagger(U)$ 
with trivial-syndrome probability $s_U=\operatorname{tr}[\Lambda_U(\rho_0)]$ and Kraus operators $ M_{\vec{j}}(U) \;\equiv\; \big\langle 0_A\big| \, U^{\dagger} K_{\vec{j}} U \, \big|0_A \big\rangle$. 
Averaging over random Clifford encoders projects this map onto the commutant of logical unitaries, yielding
\begin{equation}
  \bar{\Lambda}(\rho) = \beta \rho + \alpha \tfrac{\mathds{1}}{d_L}\operatorname{tr}(\rho), 
  \qquad d_L=2^k.
\end{equation}
After normalization by $\bar s=\alpha+\beta$, the effective normalized channel is depolarizing:
\begin{equation}
  \bar{\Lambda}^{\mathrm{cond}}(\rho) = \tau \rho + (1-\tau)\tfrac{\mathds{1}}{d_L},
  \quad \tau = \tfrac{\beta}{\alpha+\beta}.
\end{equation}
The parameter $\tau$ is fixed by the analytically computed decoding fidelity (Eq.~\ref{Eq:Taumt}), so that the effective channel behaves asymptotically as the identity channel for $p<p_c$ and reduces to complete depolarization for $p>p_c$.

\begin{figure}[h!]
    \centering
    \includegraphics[width=0.9\linewidth]{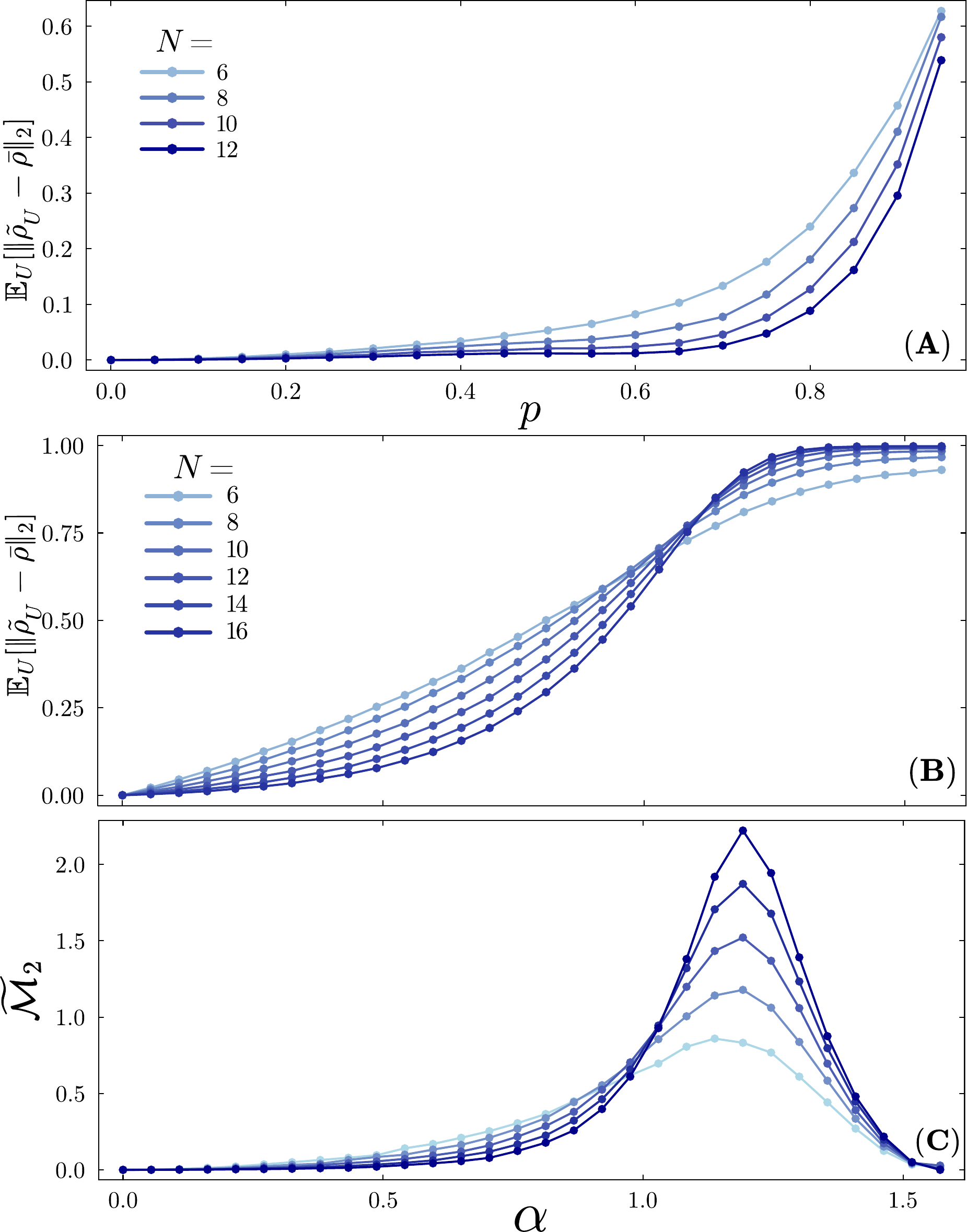}
    \caption{Hilbert–Schmidt concentration of post-selected logical outputs. 
    For amplitude damping (\textbf{A}), $\mathbb{E}_U[\|\tilde\rho_U-\bar\rho\|_2]$ decays with system size, indicating ensemble concentration onto the depolarizing channel. 
    For coherent errors (\textbf{B-C}), defined as global $Z$-rotations of angle $\alpha$ with post-selection, the HS distance fails to decay above a critical $\alpha$, where the magic monotone $\widetilde{\mathcal{M}}_2$ shows a transition.}
    \label{fig:Bound}
\end{figure}

A key diagnostic is whether the ensemble of post-selected outputs $\tilde\rho_U=\Lambda_U(\rho_0)/s_U$ concentrates around this depolarizing average $\bar\rho=\bar{\Lambda}^{\mathrm{cond}}(\rho_L)$. 
We monitor this via the average Hilbert–Schmidt (HS) distance $\mathbb{E}_U[\|\tilde\rho_U-\bar\rho\|_2]$, which is more sensitive at the accessible system sizes than trace-norm distance (the SM shows both metrics give consistent results). 
Figure (\ref{fig:Bound} \textbf{A}-\textbf{B}) displays this diagnostic as a function of system size for both AD and coherent noise. 
For AD we observe a clear decay of the HS norm for all values of $p$, signaling concentration of $\tilde\rho_U$ onto the average state $\bar \rho$; ensemble fluctuations that could carry magic are washed out with increasing $N$. In contrast, for coherent errors—implemented as a uniform $Z$-rotation by angle $\alpha$ on every qubit followed by decoding and post-selection—concentration fails: below a critical $\alpha$ the HS distance decays, while above it the distance saturates, signaling that ensemble fluctuations persist. 
Consistently, the magic monotone $\widetilde{\mathcal{M}}_2$ exhibits a transition in this coherent case (Fig.\,\ref{fig:Bound}\textbf{C}), absent for AD.

Thus, in the AD setting, the effective logical channel becomes depolarizing in the thermodynamic limit, and the robustness of magic shows no transition despite local magic generation (Fig.\,\ref{fig:Two}). For coherent noise, the absence of concentration allows a genuine nonstabilizerness transition to survive. A detailed derivation of the concentration bound, together with additional numerical evidence, is provided in the SM.\\

\begin{figure}[h!]
    \centering
    \includegraphics[width=\linewidth]{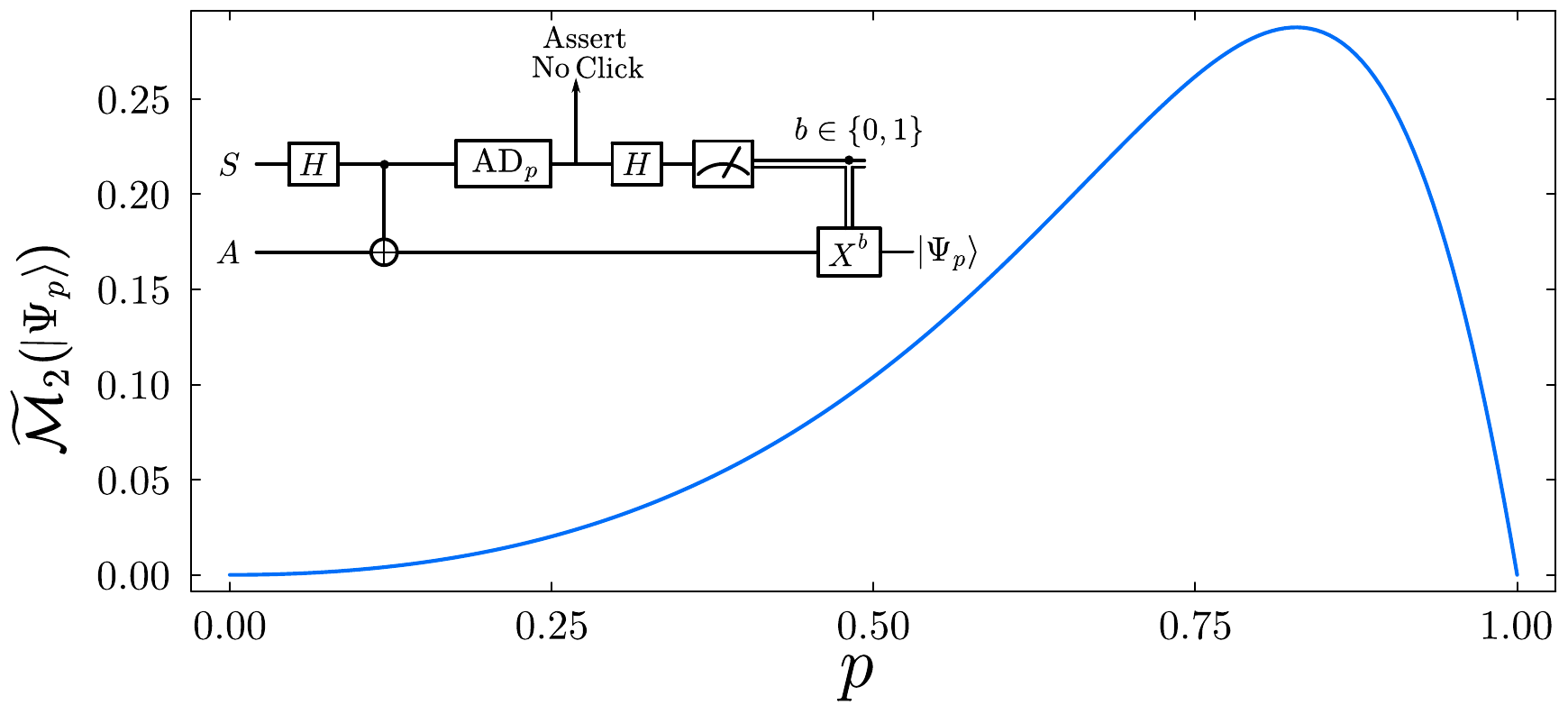}
    \caption{Stabilizer Rényi entropy of the state $|\Psi_p\rangle$ produced by the no-click amplitude damping protocol. The inset shows a schematic of the protocol.}
    \label{fig:Purif_Distill}
\end{figure}

\noindent
\textit{Magic pure states through noise.—}We describe a minimal protocol that probabilistically generates a pure magic state using only stabilizer operations and amplitude damping noise (see inset of Fig.~\ref{fig:Purif_Distill}). The protocol involves a system qubit $S$ and an ancilla $A$, initialized in $\ket{00}_{SA}$. Applying a Hadamard on $S$ followed by a CNOT prepares the Bell state $\ket{\Phi^+}_{SA}$.

We then subject the system qubit to amplitude damping with strength $p$ and condition on the no-click trajectory (no decay event). This occurs with probability $1 - p/2$ and results in the normalized state
\begin{equation}
    |\phi_0\rangle_{SA} = \frac{1}{\sqrt{2-p}}\Big( |00\rangle_{SA} + \sqrt{1-p}\, |11\rangle_{SA} \Big).
\end{equation}

Next, we measure $S$ in the $X$ basis. This projects the ancilla onto one of two pure states that differ only by a Pauli-$Z$ phase. After a conditional correction, both outcomes yield the same final state,
\begin{equation}
    \ket{\Psi_p} = \frac{\ket{0} + \sqrt{1-p}\,\ket{1}}{\sqrt{2-p}}.
\end{equation}

As shown in Fig.~\ref{fig:Purif_Distill}, $\ket{\Psi_p}$ exhibits nonzero magic for all $p \neq 0,1$. Geometrically, this family interpolates along a great circle of the Bloch sphere from $\ket{+}$ to $\ket{0}$.

\clearpage
\bibliography{NS_TO}

\onecolumngrid

\newpage
\appendix
\renewcommand{\thesection}{Appendix~\Alph{section}}

\section{Stabilizer R\'enyi Entropy of Depolarizing Noise}
\label{Ap: Analytic} 

\begin{figure}[h!]
    \centering
    \includegraphics[width=0.9\linewidth]{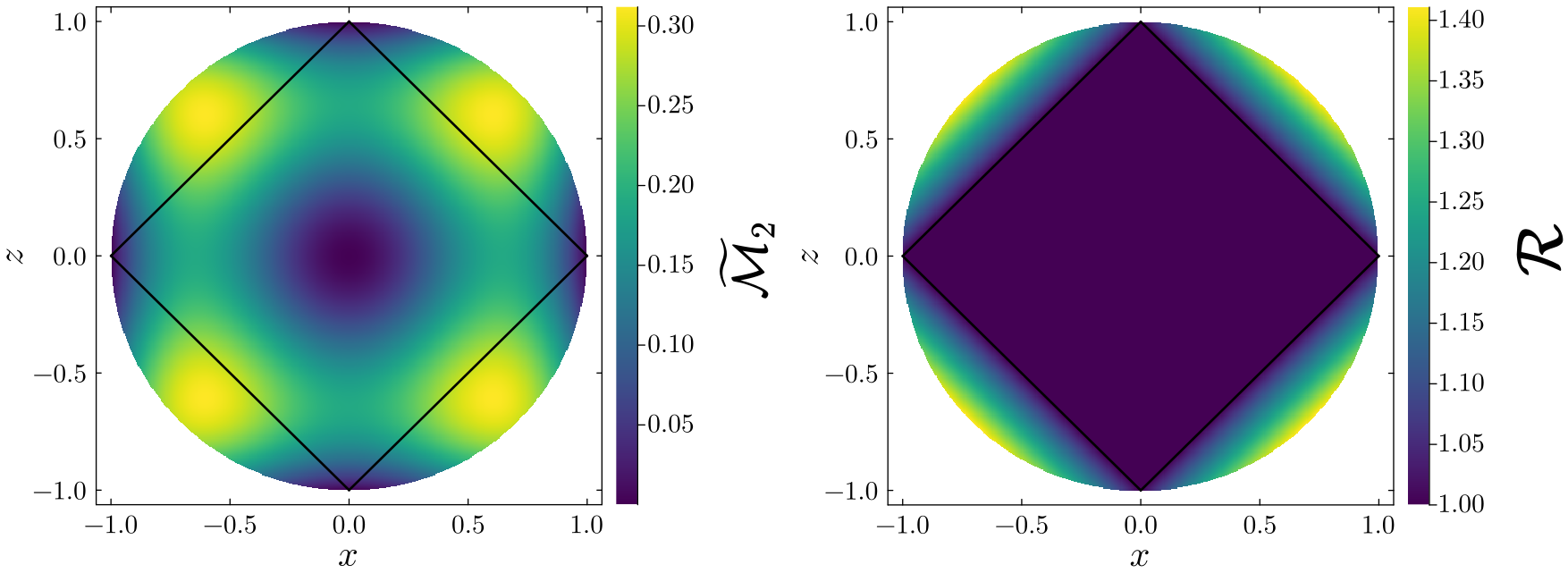}
    \caption{Heatmaps of $\widetilde{\mathcal{M}}_2$ (left) and $\mathcal{R}$ (right) on the Bloch sphere, projected onto the $xz$-plane. The black diamond outlines the stabilizer polytope. This comparison highlights the limitation of using the SRE as a measure of mixed-state magic: $\widetilde{\mathcal{M}}_2$ assigns nonzero values within the stabilizer region, whereas the ROM correctly identifies the stabilizer states as having no magic.}
    \label{fig:M2-Blochheat}
\end{figure}

To further motivate our choice of ROM as a magic measure, in this appendix, we provide an analytic computation of the SRE under the effect of depolarizing noise. It is important to emphasize that the SRE is ill-conditioned in this context and should not be used as a reliable measure. As discussed in the main text, robustness of magic is required to characterize nonstabilizerness in such cases accurately. The depolarizing noise is defined as:

\begin{align}
    \mathcal{D}_p(\rho) = \frac{p}{2^N}\mathds{1} + (1-p)\rho,    
\end{align}

\noindent
where the parameter $p$ determines the proportion of the original state $\rho$ mixed with the maximally mixed state $\mathds{1}$. Our goal is to study the effect of depolarizing noise on an initially pure stabilizer state $\rho$. Specifically, we compute the SRE extension for mixed states without the convex roof construction, as defined in Eq.~\ref{eq:SRE}. The main term to evaluate is $\Tr[\mathcal{D}_p(\rho) P]$, which can be expressed as:

\begin{equation}
    \Tr[\mathcal{D}_p(\rho) P] = \Tr\left[ \frac{p}{2^N} P + (1-p) \rho P \right] = (1-p) \Tr[\rho P] + \frac{p}{2^N} \Tr[P].
\end{equation}

\noindent
For all Pauli strings $P \neq \mathds{1}$, the last term is zero. Substituting this result, the SRE becomes:

\begin{equation}
    \overset{\sim}{\mathcal{M}}_2\big(\mathcal{D}_p(\rho)\big) = - \ln \left[ \frac{1 + (1-p)^4 \sum_{P \neq \mathds{1}} \Tr[\rho P]^4}{1 + (1-p)^2 \sum_{P \neq \mathds{1}} \Tr[\rho P]^2} \right].
\end{equation}

The structure of this expression is similar to that of the SRE for the original state, but the Pauli terms are scaled by a factor of $(1-p)$. For stabilizer states, it holds that:

\begin{align}
    \sum_{P \neq \mathds{1}} \Tr[\rho P]^2 = 2^N - 1,
\end{align}

\noindent
which follows from writing $\rho$ in the Pauli basis, $\rho = \frac{1}{2^N} \sum_{i} \alpha_i P_i$, and enforcing the purity condition. Substituting this property into the SRE expression, we arrive at:

\begin{align}
    \overset{\sim}{\mathcal{M}}_2\big(\mathcal{D}_p(\rho)\big) = -\ln \left(\frac{1 + (1-p)^4 (2^N - 1)}{1 + (1-p)^2 (2^N - 1)}\right).
    \label{eq:DepolSRE}
\end{align}

\noindent
We can further compute this quantity in the thermodynamic limit ($N \rightarrow \infty$), however, this is not more informative. This result implies that $\overset{\sim}{\mathcal{M}}_2\big(\mathcal{D}_0(\rho)\big) = \overset{\sim}{\mathcal{M}}_2\big(\mathcal{D}_1(\rho)\big) = 0$, and $\overset{\sim}{\mathcal{M}}_2\big(\mathcal{D}_p(\rho)\big) > 0$ for all $0 < p < 1$. As highlighted in the main text, this behavior contrasts sharply with the nonstabilizerness measured by the robustness of magic. This establishes that SREs (without the convex roof construction) are unsuitable as a measure of magic for mixed states.\\

An important remark: we focus on the behavior of individual states as a function of channel noise. Alternative perspectives exist, such as evaluating the general magic-generating capacity of a channel. These approaches typically require constructing the channel's associated Choi operator and computing its magic. In our study, we numerically analyzed systems of up to 4 qubits and found no evidence that the depolarizing channel generates magic. Specifically, we evaluated all possible pure stabilizer states for up to 4 qubits, applied the depolarizing channel, and tracked the robustness of magic. Across all cases, the ROM remained stable at 1, indicating that no magic was generated at any noise strength.\\

A broader argument suggests that depolarizing noise is incapable of generating magic for an initially stabilizer state. In particular, if we consider a stabilizer state, with optimal pseudomixture $\rho = \sum_i x_i |\psi_i\rangle\langle\psi_i|$, where $ x_i \geq 0 $ and $ \sum_i x_i = 1 $, and where $ |\psi_i\rangle\langle\psi_i| $ corresponds to the vertices of the stabilizer polytope. The action of the depolarizing channel can be understood geometrically as a straight-line trajectory from the initial state towards the maximally mixed state. Hence, to show that depolarizing noise cannot generate magic, it is sufficient to show that the straight line between any point inside the stabilizer polytope and the origin remains inside the polytope.\\

Rewriting the depolarizing channel in the basis of pure stabilizer states:

\begin{equation}
    \mathcal{D}_p(\rho) = (1 - p) \rho + \frac{p}{2^N} I.
\end{equation}

Using the fact that the identity can be expressed as a uniform mixture of all pure stabilizer states,

\begin{equation}
    \frac{I}{2^N} = \frac{1}{|\mathcal{S}_N|} \sum_{i} |\psi_i\rangle\langle\psi_i|,
\end{equation}

where $ |\mathcal{S}_N| $ is the number of pure $ N $-qubit stabilizer states, we obtain:

\begin{equation}
    \mathcal{D}_p(\rho) = \sum_{i} \underbrace{\left[ (1 - p) x_i + \frac{p}{|\mathcal{S}_N|} \right]}_{\equiv z_i} |\psi_i\rangle\langle\psi_i|.
\end{equation}

To show that the transformed state remains within the stabilizer polytope, we verify that the new coefficients $ z_i $ define a valid convex combination.

First, nonnegativity follows from the fact that each term in $ z_i $ is nonnegative:

\begin{equation}
    z_i = (1 - p) x_i + \frac{p}{|\mathcal{S}_N|} \geq 0.
\end{equation}

Next, normalization is verified as follows:

\begin{equation}
    \sum_i z_i = (1 - p) \underbrace{\sum_i x_i}_{=1} + \frac{p}{|\mathcal{S}_N|} \underbrace{\sum_i}_{|\mathcal{S}_N|}.
\end{equation}

Since $ \sum_i x_i = 1 $ and $ \sum_i 1 = |\mathcal{S}_N| $, we obtain:

\begin{equation}
    \sum_i z_i = (1 - p) + p = 1.
\end{equation}

Since $ \{z_i\} $ define a valid convex combination of stabilizer states, the resulting state remains inside the stabilizer polytope. 

Thus, we conclude that depolarizing noise cannot generate magic from an initially stabilizer state. The depolarizing channel simply shrinks the Bloch vector toward the maximally mixed state, and since the stabilizer polytope is convex, this trajectory remains inside it at all times.

\section{\label{Ap: AD_geometry}Amplitude Damping Trajectories and the Geometry of Nonstabilizerness}

\begin{figure}[h!]
    \centering
    \includegraphics[width=0.35\linewidth]{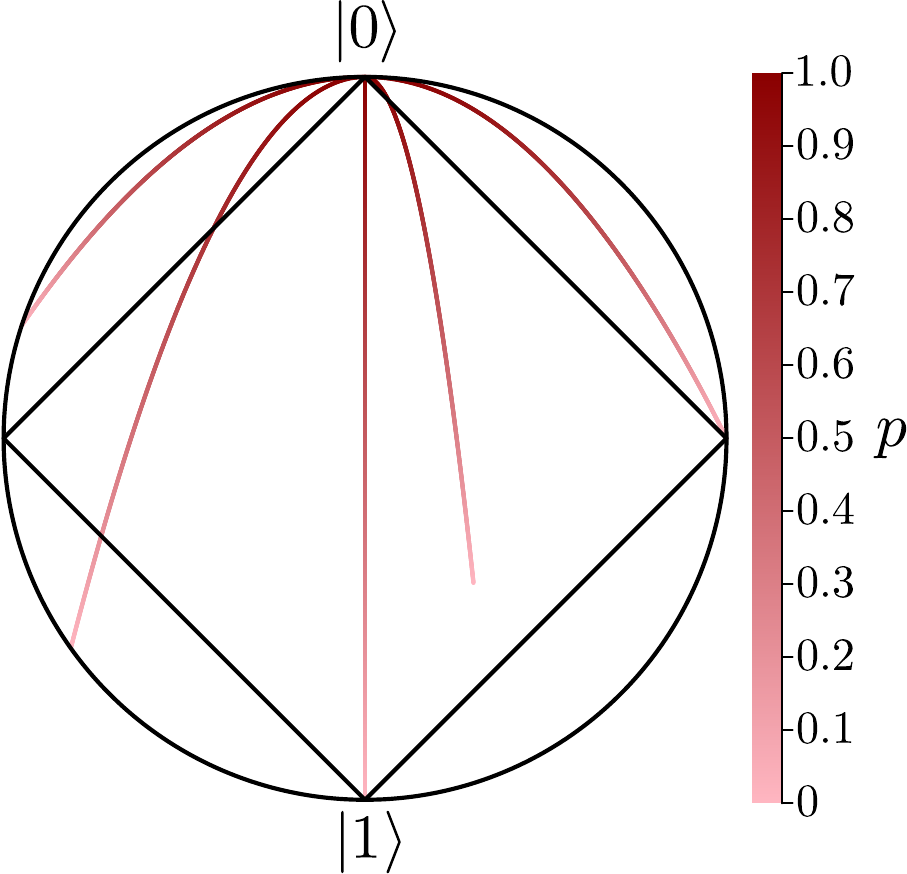}
    \caption{Bloch sphere trajectories of certain states under amplitude damping noise.}
    \label{fig:ad_gEO}
\end{figure}

As we mentioned in the main text, magic generation can be geometrically visualized in a clean manner for a single qubit undergoing amplitude-damping noise. In Fig. \ref{fig:ad_gEO}, we show a cut of the Bloch sphere through the $y\text{-}z$ plane. The stabilizer polytope region is pictured by the squared section therein. $p$ tunes trajectories of arbitrary points (initial states) to the north pole. As seen in the figure, these trajectories can explore regions outside of the stabilizer polytope for initial stabilizer states, which injects magic into the system. 

\section{\label{Ap: GADC} Generalization of Amplitude Damping Noise}

In this section, we present a generalization to the amplitude damping noise discussed in the main text. The generalized amplitude-damping channel (GADC) is a noise model that describes the influence of an environment at a finite temperature on a quantum state \cite{Nielsen10, Khatri20}. The single-qubit Kraus operators that define the GADC are 

\begin{equation}
\begin{aligned}
K_0 &= \sqrt{\eta} \begin{pmatrix} 1 & 0 \\ 0 & \sqrt{1{-}p} \end{pmatrix},\quad 
K_1 = \sqrt{\eta} \begin{pmatrix} 0 & \sqrt{p} \\ 0 & 0 \end{pmatrix}, \\
K_2 &= \sqrt{1{-}\eta} \begin{pmatrix} \sqrt{1{-}p} & 0 \\ 0 & 1 \end{pmatrix},\quad 
K_3 = \sqrt{1{-}\eta} \begin{pmatrix} 0 & 0 \\ \sqrt{p} & 0 \end{pmatrix}.
\end{aligned}
\label{Eq:Kraus}
\end{equation}

We denote the GADC's action on a state $\rho$ as $\mathcal{G}_{p,\eta}(\rho) = \sum_{i=0}^3 K_i \rho K_i^{\dagger}$.
A qubit state can generally be written in its Bloch form: $\rho = \frac{1}{2}(\mathds{1}+r_x X+r_y Y+r_z Z)$, where \textbf{r} $=(r_x,r_y,r_z)$ is the Bloch vector. The GADC's action on a state  changes the state's Bloch vector:
\begin{equation}
    \textbf{r}' = \left(r_x \sqrt{1 - p},\, r_y \sqrt{1 - p},\, p(2\eta-1) + r_z (1 - p)  \right).
\end{equation}

The GADC collapses the Bloch sphere to an ellipsoid with its center on the positive $z$ axis \footnote{One can generalize the form of GADC we take here to a channel that collapses the Bloch sphere to any arbitrary point by simply taking an appropriate rotation of the $z$ axis.}. Through reparametrizations, one can interpret the GADC's knobs as indicators of a system's relaxation time scale \cite{Nielsen10} or or view the GADC as a qubit thermal channel \cite{Khatri20}. 

$p$ represents a damping parameter, while $\eta$ encodes the environment's temperature \cite{Cafaro14}: $\eta = 1-\frac{e^{-\omega/T}}{1+e^{-\omega/T}}$, where $\omega$ denotes the qubit's energy gap. The standard amplitude-damping channel corresponds to $ \eta = 1 $, describing zero-temperature relaxation. In this limit, the dynamics drive any state toward the Bloch sphere's north pole, $ |0\rangle $. In the main text, we denote this channel by $\mathcal{N}_p(\rho) = \mathcal{G}_{p,1}(\rho) $.

\begin{figure}[h!]
    \centering
    \includegraphics[width=0.61\linewidth]{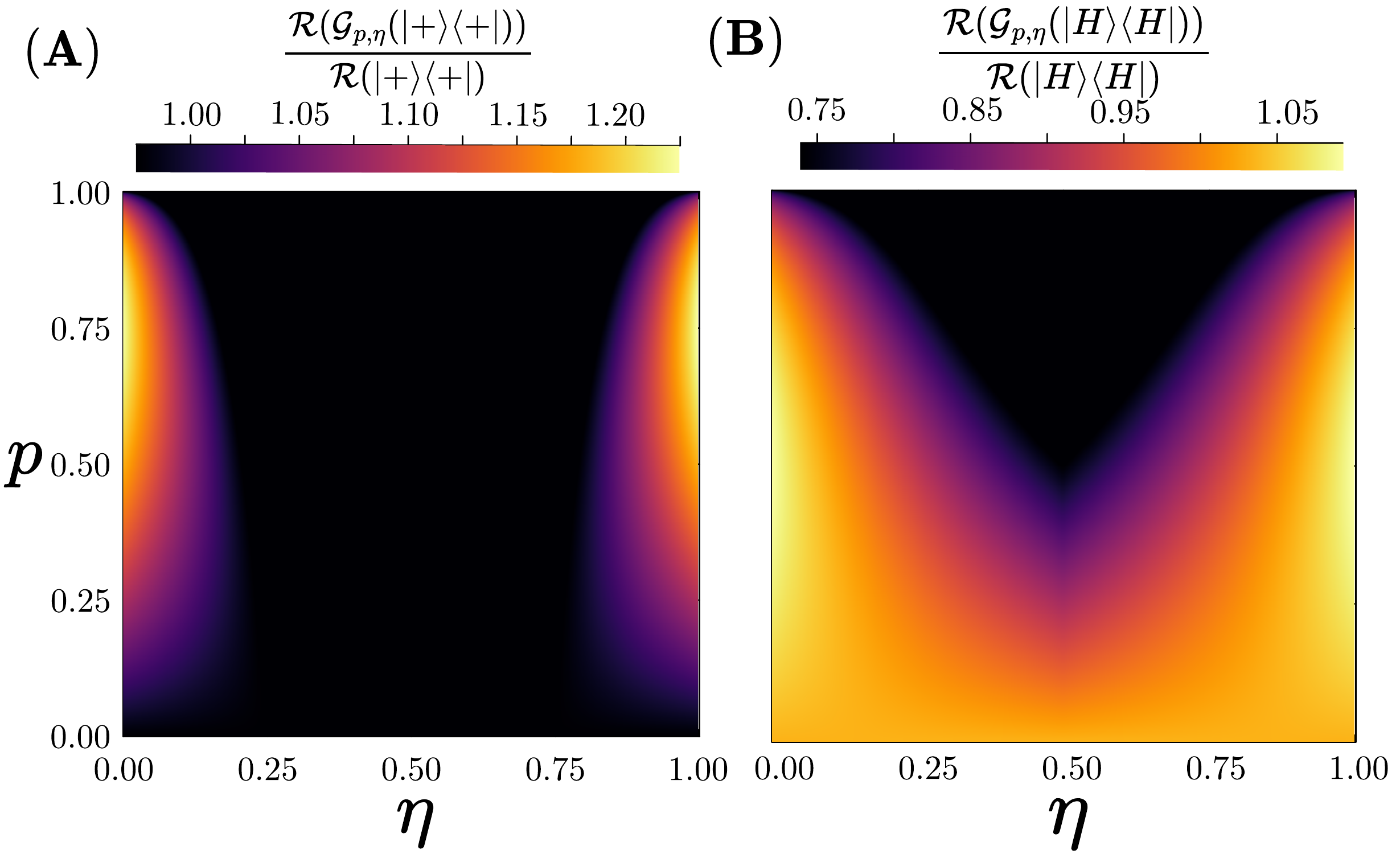}
    \caption{ Heat maps of ROM for the generalized amplitude damping channel applied to $|+\rangle$ and $|H\rangle$ across different channel parameters.}
    \label{fig:GADC}
\end{figure}

Figure~\ref{fig:GADC} shows the ratio of the ROM of a state after undergoing the GADC to the ROM of the original state. Panels (A) and (B) correspond to the stabilizer state $|+\rangle$ and the magic state $|H\rangle$, respectively, as the parameters $p$ and $\eta$ are varied. Magic persists at finite temperature: for small $\eta$, the channel generates magic, and for $|H\rangle$, ROM remains large across a broad parameter range. Only when both $p$ and $\eta$ are large is the resource completely suppressed. Thus, the coupling of a quantum system to a thermal reservoir can both generate and sustain magic.

\section{\label{Ap: Fidelity} Fidelity Averaged Over the Haar Measure for Amplitude Damping}

To analytically compute the Haar-averaged fidelity in the presence of amplitude damping, we follow the approach outlined in \cite{PhysRevLett.132.140401}. A detailed derivation can be found in that work; here, we provide a high-level overview of the key steps. Notably, since the Clifford group forms a unitary 3-design \cite{zhu2016cliffordgroupfailsgracefully,cliffordcommutant}, the same analytic approach also yields the fidelity averaged over Clifford unitaries, as both averages coincide.

The computation relies on the $k$-replica trick, where the principal object is the channel average:

\begin{equation}
    \Phi_{\text{Haar}}^{(k)}(\mathcal{O}) = \int_{U \in \mathcal{U}(2^N)} d\mu(U) \left[ \left(U^{\dagger}\right)^{\otimes k} \mathcal{O} \ U^{\otimes k} \right].
\end{equation}

Since $\Phi_{\text{Haar}}^{(k)}(\mathcal{O})$ belongs to the commutant of $\mathcal{U}(2^N)$, Schur-Weyl duality allows us to express it as a linear combination of permutation operators, significantly simplifying the computation \cite{Gross_2021}:

\begin{equation}
    \Phi_{\text{Haar}}^{(k)}(\mathcal{O}) = \sum_{\pi \in S_k} b_{\pi}(\mathcal{O}) T_{\pi}.
\end{equation}

\begin{figure}[h!]
    \centering
    \includegraphics[width=0.45\linewidth]{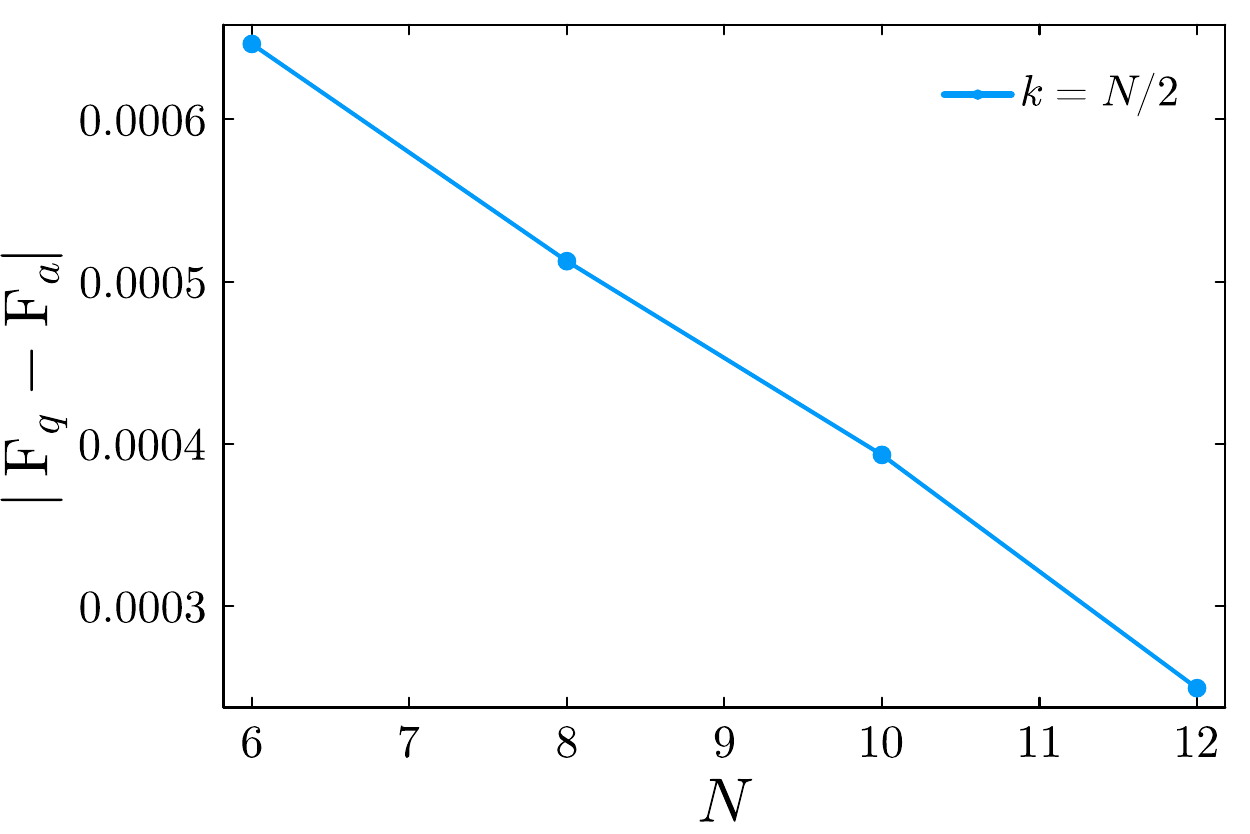}
    \caption{Difference between the quenched and annealed averages at code rate $r = N/2$ as a function of system size.}
    \label{fig:QADIF}
\end{figure}

Here, $T_{\pi}$ are representations of the permutation group $S_k$ on the $k$-replica space, and the coefficients are given by

\begin{equation}
    b_{\pi}(\mathcal{O}) = \sum_{\sigma \in S_k} W_{\pi, \sigma} \Tr(\mathcal{O} T_{\sigma}),
\end{equation}

where $W_{\pi, \sigma}$ are the Weingarten symbols \cite{Roberts_2017}. As shown in \cite{PhysRevLett.132.140401}, the fidelity in our encoding-decoding setup can be computed via the replica formalism as $\Phi^{(2)}(F)$, where

\begin{equation}
    F = \frac{\langle 0_{\bar{X}}|U^{\dagger} \mathcal{N}_p(U \rho_0 U^{\dagger}) U|0_{\bar{X}}\rangle}{\Tr[\langle 0_{\bar{X}}|U^{\dagger} \mathcal{N}_p(U \rho_0 U^{\dagger}) U|0_{\bar{X}}\rangle]},
\end{equation}

with $\rho_0 = |0\rangle\langle0|^{\otimes N}$ as the initial state and $\mathcal{N}_p$ the amplitude damping channel. An important remark is that here, we are technically averaging the numerator and denominator separately, and there is no guarantee that the ratio might have self-averaging. That is why in our data, we also show finite-size numerical data to show how this matches. Furthermore, in Fig.\,(\ref{fig:QADIF}) we present numerical data of the difference between the quenched and annealed averages, respectively defined as:
\begin{equation*}
F_q = \mathbb{E}_U\left[\frac{\langle 0_{\bar{X}}|U^{\dagger} \mathcal{N}_p(U \rho_0 U^{\dagger}) U|0_{\bar{X}}\rangle}{\Tr[\langle 0_{\bar{X}}|U^{\dagger} \mathcal{N}_p(U \rho_0 U^{\dagger}) U|0_{\bar{X}}\rangle]}\right] \quad\text{and } \quad F_a = \frac{\mathbb{E}_U[\langle 0_{\bar{X}}|U^{\dagger} \mathcal{N}_p(U \rho_0 U^{\dagger}) U|0_{\bar{X}}\rangle]}{\mathbb{E}_U[\Tr[\langle 0_{\bar{X}}|U^{\dagger} \mathcal{N}_p(U \rho_0 U^{\dagger}) U|0_{\bar{X}}\rangle]}.
    \label{eq:QuenchedAnnealed}
\end{equation*}

The numerical data shows that the values are very close at the system sizes we consider and that the trend seems to decay as system size increases. Hence, we can take analytics to represent the behavior of the post-selected state with high confidence that the behavior is representative of the correct physics.

The problem thus reduces to computing $b_{\pi}$ for amplitude damping, where the Kraus operators are given in Eq.~(\ref{Eq:Kraus}):

\begin{equation}
    b_{\pi}(\mathcal{N}_p) = \sum_{\tau} W_{\pi, \tau} \Tr\left[ t_{\pi} \left( \sum_{i=0}^{1} K_{i} \otimes K_{i}^{\dagger} \right)^N \right].
\end{equation}

The term inside the trace, denoted as $Q$, takes the following form for amplitude damping:

\begin{equation}
    Q = \begin{bmatrix}
    1 & 0 & 0 & 0 \\
    0 & \sqrt{1 - p} & p & 0 \\
    0 & 0 & \sqrt{1 - p} & 0 \\
    0 & 0 & 0 & 1 - p
    \end{bmatrix}.
\end{equation}

With this, the fidelity simplifies to

\begin{equation}
    F_{\textrm{AD}} = \frac{(2^N - 1) \left[ 1 + (1 + \sqrt{1-p} - p/2)^N \right]}{(2^N - 2^k)\left[ 1 + \sqrt{1-p} - p/2 \right]^N + 2^{N+k} - 1}.
\end{equation}

Taking the limit $N \to \infty$, we observe a phase transition between an error-protecting and an error-vulnerable phase at $p_c = \left( -2^{2+r} + 2^{(5+r)/2} \right)/2$.

\section{\label{Ap: SRE-Enc} Encoding-Decoding Transition, SREs and Witnesses}

In this section, we present the numerical results of computing SRE-based measures ($\widetilde{\mathcal{M}}_2$ and  $\mathcal{W}_2(\rho)$) as probes for the nonstabilizerness transition in the encoding-decoding setup. Therefore, this plot is analogous to that in Fig.\,\ref{fig:One}(\textbf{C}), but we use different measures of magic to track the potential transition. 

\begin{figure}[h!]
    \centering
    \includegraphics[width=0.5\linewidth]{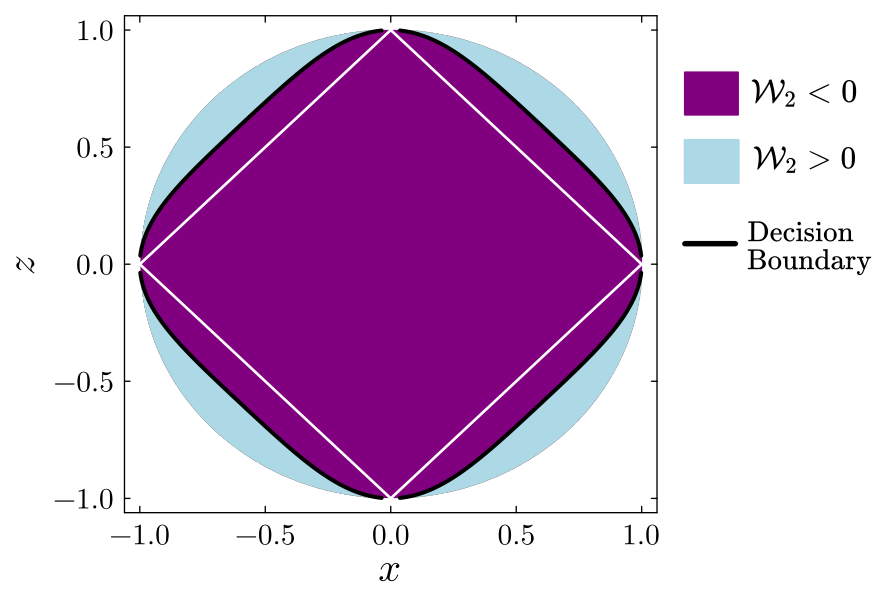}
    \caption{Bloch sphere cross section with the decision boundary for the magic witness $\mathcal{W}_2. $}
    \label{fig:W2_deci}
\end{figure}

$\mathcal{W}$ is a new measure based on the SREs for mixed states has recently been proposed as a magic witness; it is not a true magic monotone \cite{MagicWitness}. This quantity is defined as
\begin{equation}
    \mathcal{W}_2(\rho) = \widetilde{\mathcal{M}}_2(\rho) - 2 S_2(\rho),
\end{equation}
where \( S_2(\rho) = -\ln\left(\mathrm{Tr}(\rho^2)\right) \) denotes the log-purity of the quantum state.
\noindent
The witness \( \mathcal{W}_2(\rho) \) indicates the presence of magic: a positive value certifies that the state is magical. However, if \( \mathcal{W}_2(\rho) \) is negative, no definitive conclusion can be drawn; the state could still possess magic that the witness fails to detect. Despite this limitation, \( \mathcal{W}_2(\rho) \) is significantly better suited for studying mixed-state magic than SREs (see Fig.\,\ref{fig:W2_deci}) and can be efficiently computed. The filtered variant of this quantity is said to be much more sensitive, which could help even further in having better diagnosis of mixed states, however we do not analyze that scenario in this work. Since it is not a true monotone and thus does not faithfully capture nonstabilizerness across general operations, we do not rely on it for our primary analysis.

\begin{figure}[H]
    \centering
    \includegraphics[width=0.9\linewidth]{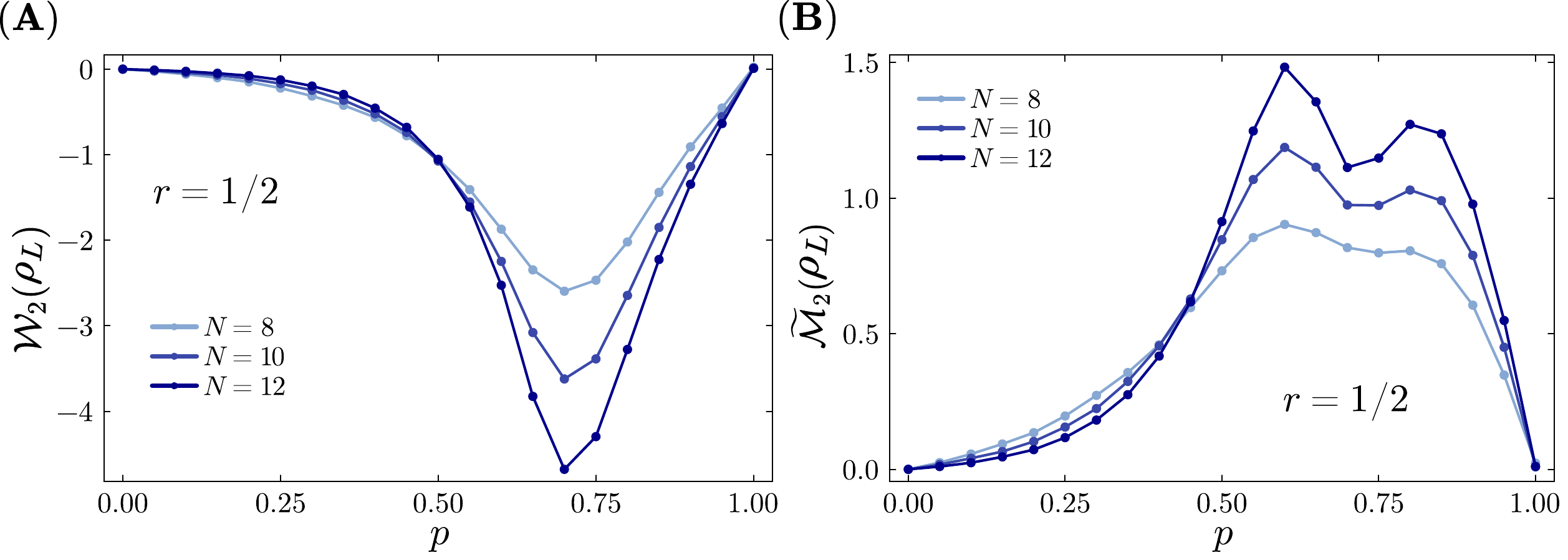}
    \caption{(\textbf{A}) Magic witness and (\textbf{B}) Stabilizer R\'enyi entropy of the encoding-decoding set up for amplitude damping noise as a function of damping parameter. These are averaged over 1000 realizations of Clifford unitaries.}
    \label{fig:SRE-tr}
\end{figure}

In Fig.\,\ref{fig:SRE-tr}(\textbf{B}), we show that SREs suggest the presence of a phase transition close to that of the encoding-decoding transition. This contrasts the observed behavior of ROM as a nonstabilizerness probe. We highlight that we believe this behavior is not indicative of a true resource transition but rather just an artifact of SREs' being bad magic monotones for mixed states. On the other hand, Fig.\,\ref{fig:SRE-tr}(\textbf{A}) shows the behavior of the magic witness $\mathcal{W}_2$ under the same protocol. The witness generally stays negative, which highlights that the measure finds no evidence of magic present in the states. However, this does not readily imply the absence of magic due to the nature of witnesses. 

\section{\label{Ap: LargeAD} Amplitude Damping Magic Generation for Larger Systems}

\begin{figure}[h!]
    \centering
    \includegraphics[width=0.5\linewidth]{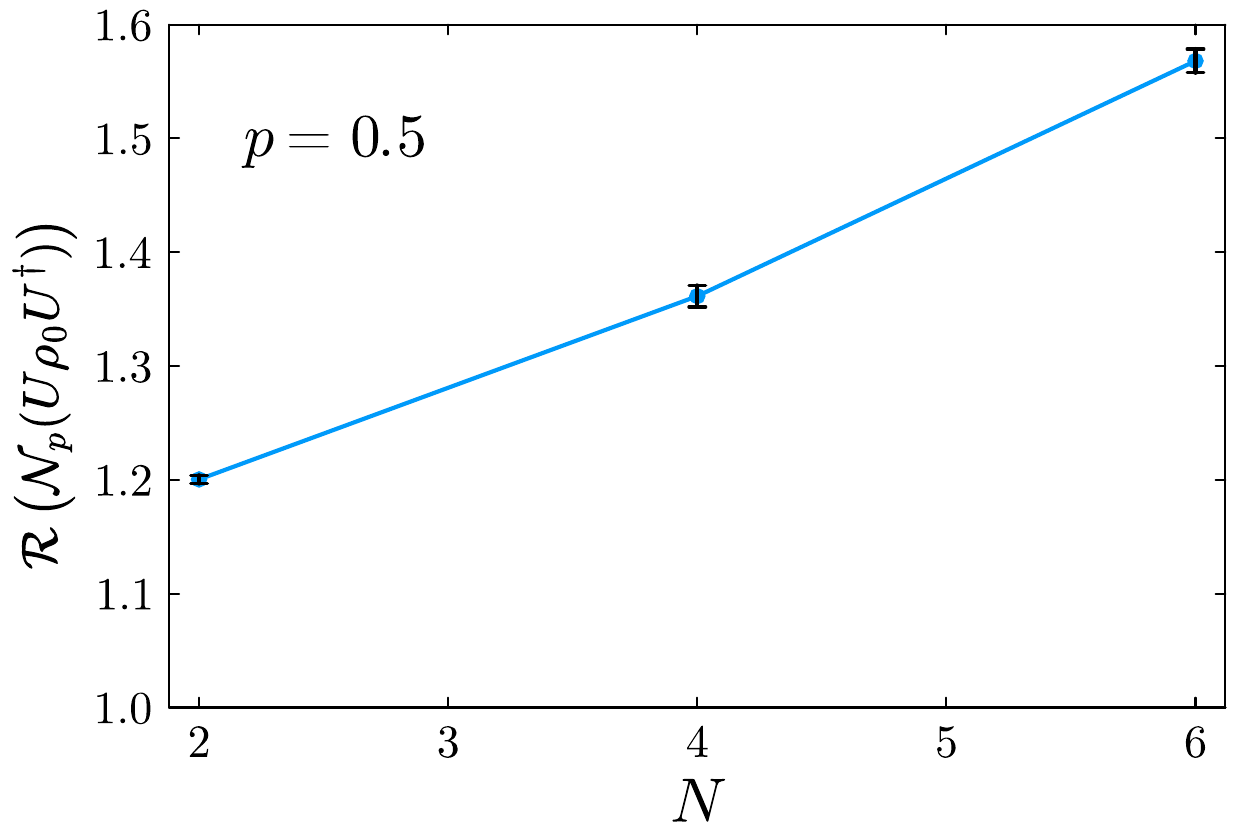}
    \caption{Robustness of magic for the states after the error layer in the encoding-decoding protocol at different system sizes. We average over $500$ realizations of random Clifford unitaries.}
    \label{fig:LargeAD}
\end{figure}

The decay observed in the ROM in Fig.\,\ref{fig:One}(\textbf{C}) raises an important question about the origin of this diminishing magic. Specifically, it is not evident whether the amplitude damping channel inherently produces less magic as the number of qubits increases—implying that no magic is generated in the error layer in the thermodynamic limit—or whether the observed suppression is instead due to the destructive effect of post-selection associated with measurement. To resolve this, Fig.\,\ref{fig:LargeAD} presents the ROM evaluated immediately after the error layer—before any measurement or decoding steps—demonstrating that the magic generated by the noise does not decrease with system size. This indicates that the suppression in Fig.\,\ref{fig:One}(\textbf{C}) arises from post-selection, not from the channel itself.

\section{\label{Ap:NoClick} Magic generation via amplitude damping and no-click post-selection}

In this appendix, we provide a detailed derivation of the protocol outlined in the end matter for generating pure magic states using amplitude damping and post-selection. The key idea is to exploit the conditional (“no-click”) trajectories of amplitude damping, which transform stabilizer inputs into nonstabilizer states.

Consider a system qubit $S$ and an ancilla $A$, both initialized in $\ket{0}$. The initial two-qubit state is
\begin{equation}
    |\phi_0\rangle_{SA} = |0\rangle_S \otimes |0\rangle_A.
\end{equation}
We apply a Hadamard gate to $S$ and then a CNOT from $S$ to $A$, preparing the Bell state
\begin{equation}
    |\Phi^+\rangle_{SA} = \tfrac{1}{\sqrt{2}}\big( |00\rangle_{SA} + |11\rangle_{SA} \big).
\end{equation}

Next, the system qubit undergoes amplitude damping with damping strength $p$. The amplitude damping channel $\mathcal{N}_p$ is defined by Kraus operators
\begin{equation}
    K_0 = \ket{0}\!\bra{0} + \sqrt{1-p}\,\ket{1}\!\bra{1}, 
    \qquad
    K_1 = \sqrt{p}\,\ket{0}\!\bra{1}.
\end{equation}
The operator $K_0$ corresponds to the “no-click” trajectory (no quantum jump), while $K_1$ corresponds to the “click” trajectory (a jump $\ket{1}\to\ket{0}$).

Applying $K_0$ to the system qubit yields the unnormalized state
\begin{equation}
    |\phi_0\rangle_{SA} = (K_0 \otimes \mathds{1}_A)|\Phi^+\rangle_{SA}
    = \tfrac{1}{\sqrt{2}} \big( |00\rangle_{SA} + \sqrt{1-p}\,|11\rangle_{SA} \big).
\end{equation}
The probability of this trajectory is
\begin{equation}
    P_{\text{no-click}} = \|(K_0 \otimes \mathds{1}_A)|\Phi^+\rangle_{SA}\|^2 = \tfrac{1}{2}(1 + 1-p) = 1 - \tfrac{p}{2}.
\end{equation}
Conditioned on this outcome, the normalized state is
\begin{equation}
    |\phi_0\rangle_{SA} = \frac{1}{\sqrt{2-p}} \big( |00\rangle_{SA} + \sqrt{1-p}\,|11\rangle_{SA} \big).
\end{equation}

We now measure the system qubit $S$ in the $X$ basis. Writing $\ket{\pm} = (\ket{0}\pm \ket{1})/\sqrt{2}$, the state becomes
\begin{equation}
    |\phi_0\rangle_{SA} = 
    \tfrac{1}{\sqrt{2-p}}\Big(
        \tfrac{1}{\sqrt{2}}|+\rangle_S(|0\rangle_A + \sqrt{1-p}\,|1\rangle_A)
        + \tfrac{1}{\sqrt{2}}|-\rangle_S(|0\rangle_A - \sqrt{1-p}\,|1\rangle_A)
    \Big).
\end{equation}
Upon measurement, outcome $\ket{+}$ leaves the ancilla in
\[
\ket{\Psi_p} = \frac{\ket{0} + \sqrt{1-p}\ket{1}}{\sqrt{2-p}},
\]
while outcome $\ket{-}$ leaves it in
\[
\ket{\Psi_p^-} = \frac{\ket{0} - \sqrt{1-p}\ket{1}}{\sqrt{2-p}}.
\]
A conditional Pauli $Z$ correction on the ancilla maps both outcomes to the same state $\ket{\Psi_p}$.

\section{\label{Ap: Effec} Effective Channel and Ensemble Averages}

To analyze why a magic transition is absent under amplitude damping, we study the effective logical channel. For a fixed encoder $U$, the (unnormalized) selective map that includes noise, decoding, and post-selection onto the trivial syndrome is
\begin{equation}
  \Lambda_U(\rho) \;=\; 
  \big\langle 0_A \big|\, U^{\dagger}\, \mathcal{N}_p\!\Big( U \,\big(\rho \otimes |0\rangle\!\langle0|_A\big)\, U^{\dagger}\Big)\, U \,\big|0_A \big\rangle.
  \label{eq:unnorm}
\end{equation}
Writing $\mathcal{N}_p$ in Kraus form with operators $\{K_{\vec{j}}\}$, define
\begin{equation}
    M_{\vec{j}}(U) \;\equiv\; \big\langle 0_A\big| \, U^{\dagger} K_{\vec{j}} U \, \big|0_A \big\rangle,
\end{equation}
so that
\begin{equation}
    \Lambda_U(\rho) \;=\; \sum_{\vec{j}} M_{\vec{j}}(U)\, \rho\, M_{\vec{j}}^{\dagger}(U).
\end{equation}
This map is completely positive but generally trace-decreasing. The probability of the trajectory's yielding a trivial syndrome is $s_U \equiv \operatorname{tr}[\Lambda_U(\rho)]$. The conditional (trace-preserving) logical channel is
\begin{equation}
  \Lambda_U^{\mathrm{cond}}(\rho) \;=\; \frac{\Lambda_U(\rho)}{s_U}.
\end{equation}

\textbf{Averaging over encoders.} Let $\bar{\Lambda} \equiv \mathbb{E}_U[\Lambda_U]$ denote the ensemble average of the unnormalized map. Because the Clifford group is a unitary $2$-design, twirling projects any channel onto the commutant of logical unitaries \cite{Nielsen_2002, Emerson_2005, horodecki1998}. The only completely positive maps with this symmetry are
\begin{equation}
  \bar{\Lambda}(\rho) \;=\; \beta\,\rho \;+\; \alpha\,\frac{\mathbb{I}}{d_L}\,\operatorname{tr}(\rho),
  \qquad d_L=2^k,
  \label{eq:twirled}
\end{equation}
with $\alpha,\beta\ge 0$. Normalizing by $\bar{s}=\operatorname{tr}[\bar{\Lambda}(\rho)]=\alpha+\beta$ yields the averaged conditional channel, which is depolarizing:
\begin{equation}
  \bar{\Lambda}^{\mathrm{cond}}(\rho)
  \;=\; \tau\,\rho \;+\; (1-\tau)\,\frac{\mathbb{I}}{d_L},
  \qquad
  \tau \equiv \frac{\beta}{\alpha+\beta}.
  \label{eq:dep}
\end{equation}

We can fix the value of $\tau$ from the expression for fidelity. For a pure input $\rho=|\psi\rangle\!\langle\psi|$, the decoding fidelity is linear in the state, hence
\begin{equation}
  F_{\mathrm{AD}} \;=\; \mathbb{E}_U\!\big[\operatorname{tr}(\Lambda_U^{\mathrm{cond}}(\rho)\,\rho)\big]
  \;=\; \operatorname{tr}\!\big(\bar{\Lambda}^{\mathrm{cond}}(\rho)\,\rho\big)
  \;=\; \tau \;+\; \frac{1-\tau}{d_L}.
\end{equation}
Solving gives
\begin{equation}
  \tau \;=\; \frac{d_L\,F_{\mathrm{AD}}-1}{d_L-1}.
\end{equation}
Using the replica expression for $F_{\mathrm{AD}}$,
\begin{equation}
  \tau(p,N,k)
  \;=\; \frac{2^{N}\,\lambda(p)^{N}-1}{\big(2^{N}-2^{k}\big)\lambda(p)^{N}+2^{N+k}-1},
  \qquad
  \lambda(p)=1+\sqrt{1-p}-\frac{p}{2}.
  \label{eq:tau}
\end{equation}

As $N\to\infty$ at fixed rate $r=k/N$, $\lambda(p)$ crosses $2^{\,r}$ at a critical $p_c$, so
\begin{equation}
  \lim_{N\to\infty}\tau(p,N,k)
  \;=\;
  \begin{cases}
    1, & p<p_c,\\
    0, & p>p_c.
  \end{cases}
\end{equation}
Equivalently,
\begin{equation}
  \lim_{N\to\infty}\bar{\Lambda}^{\mathrm{cond}}(\rho)
  \;=\;
  \begin{cases}
    \rho, & p<p_c,\\[2pt]
    \mathbb{I}/d_L, & p>p_c.
  \end{cases}
\end{equation}

\begin{figure}[t]
    \centering
    \includegraphics[width=0.45\linewidth]{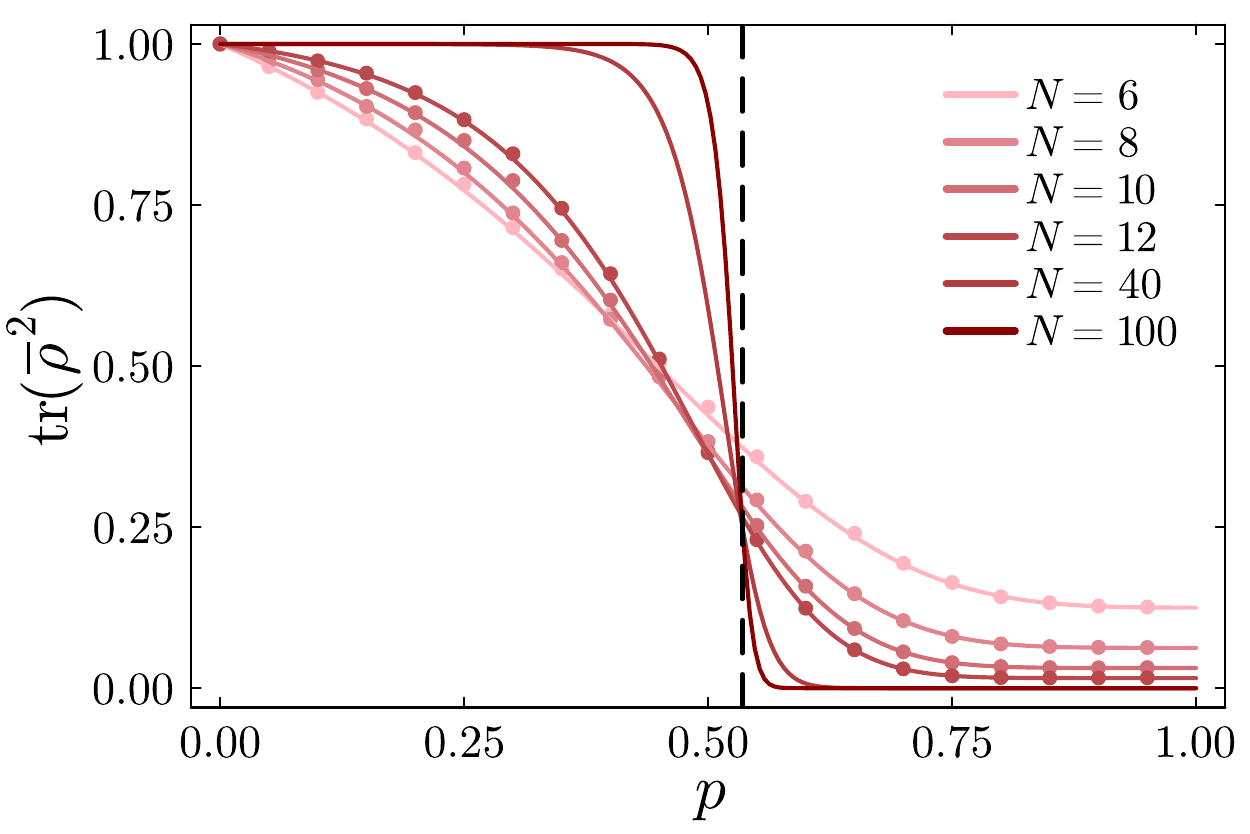}
    \caption{Purity of the averaged (post-selected) logical state $\bar\rho$ vs AD strength $p$, for various $N$ (markers: numerics; lines: Eq.~\eqref{eq:purity-mean}). The vertical line marks $p_c$.}
    \label{fig:MeanPuri}
\end{figure}

We now look at the mean state's purity.  For any logical input,
\begin{equation}
  \operatorname{tr}(\bar{\rho}^2)
  \;=\; \tau^2\,\operatorname{tr}(\rho_L^2) \;+\; \frac{1-\tau^2}{d_L},
  \qquad \bar{\rho} \equiv \bar{\Lambda}^{\mathrm{cond}}(\rho_L).
  \label{eq:purity-mean}
\end{equation}
In Fig.~\ref{fig:MeanPuri} we plot \eqref{eq:purity-mean} against numerics, showing the crossover: $\operatorname{tr}(\bar{\rho}^2)\!\to\!1$ for $p<p_c$ and $\operatorname{tr}(\bar{\rho}^2)\!\to\!1/d_L$ for $p>p_c$.

\textbf{Ensemble averages and concentration.} Here, we expand on the end matter's discussion on concentration around the mean. Let $\tilde{\rho}_U \equiv \Lambda_U^{\mathrm{cond}}(\rho_0)$ be the normalized post-selected output and $\bar\rho = \mathbb{E}_U[\tilde{\rho}_U]$ the ensemble mean.
Since the robustness of magic is a generalized robustness over the stabilizer polytope, it is Lipschitz in trace norm \cite{Schluck_2023}:
\begin{equation}
  \big|\mathcal{R}(\rho)-\mathcal{R}(\sigma)\big|
  \;\le\; L(d_L)\,\|\rho-\sigma\|_1,
\end{equation}
for some $L(d_L)$ depending only on the logical dimension $d_L=2^k$. Therefore
\begin{equation}
  0 \;\le\; \mathbb{E}_U\!\big[\mathcal{R}(\tilde{\rho}_U)\big] - \mathcal{R}(\bar{\rho})
  \;\le\; L(d_L)\,\mathbb{E}_U\!\big[\|\tilde{\rho}_U-\bar{\rho}\|_1\big].
  \label{eq:ROM-gap-again}
\end{equation}
It remains to control $\mathbb{E}_U\|\tilde{\rho}_U-\bar{\rho}\|_1$.

\begin{figure}[h!]
    \centering
    \includegraphics[width=\linewidth]{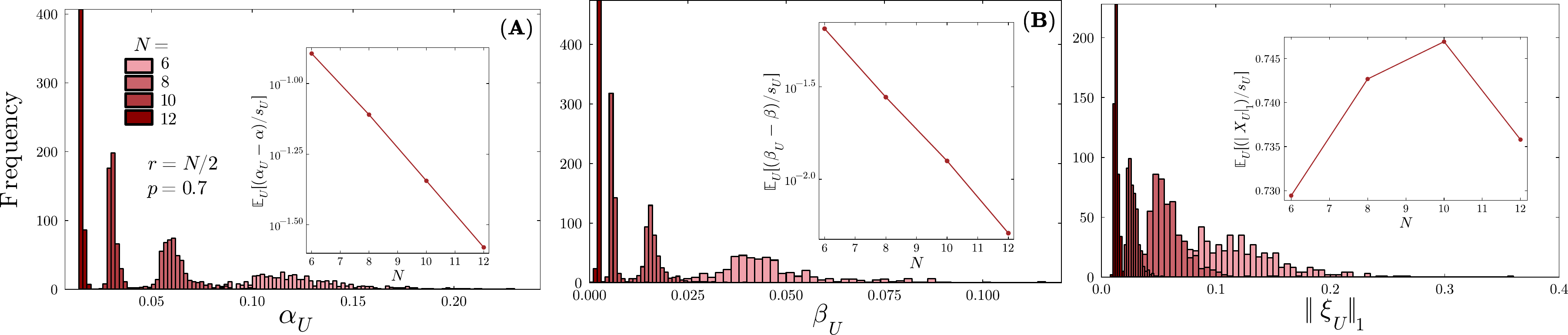}
    \caption{(\textbf{A}) Histogram of $\alpha_U$ with inset showing $\mathbb{E}_U[(\alpha_U-\alpha)/s_U]$ versus $N$.  
    (\textbf{B}) Histogram of $\beta_U$ with inset showing $\mathbb{E}_U[(\beta_U-\beta)/s_U]$ versus $N$.  
    (\textbf{C}) Histogram of $\|\xi_U\|_1$ with inset showing $\mathbb{E}_U[\|\xi_U\|_1/s_U]$ versus $N$.}
    \label{fig:alphabetaXi}
\end{figure}

Fix a pure logical input $\rho_L$ and write the unnormalized selective output as
\begin{equation}
  \Lambda_U(\rho_0) \;=\; \alpha_U\,\frac{\mathbb{I}}{d_L} \;+\; \beta_U\,\rho_L \;+\; \xi_U,
  \qquad \operatorname{tr}(\xi_U)=0,\quad \operatorname{tr}(\rho_L\xi_U)=0,
  \label{eq:orth-split}
\end{equation}
where $\alpha_U,\beta_U$ are determined by
\begin{equation}
  s_U\equiv\operatorname{tr}\Lambda_U(\rho_0)=\alpha_U+\beta_U,\qquad
  t_U\equiv\operatorname{tr}(\rho_L\Lambda_U(\rho_0))=\frac{\alpha_U}{d_L}+\beta_U,
\end{equation}
that is,
\begin{equation}
  \alpha_U=\frac{d_L}{d_L-1}(s_U-t_U),\qquad
  \beta_U=\frac{d_L\,t_U-s_U}{d_L-1}.
\end{equation}
Taking ensemble means $\alpha=\mathbb{E}[\alpha_U]$, $\beta=\mathbb{E}[\beta_U]$ gives the averaged unnormalized map $\bar\Lambda(\rho_L)=\alpha\,\frac{\mathbb{I}}{d_L}+\beta\,\rho_L$, and the averaged conditional state $\bar\rho=\bar\Lambda/\bar s$ where $\bar s=\alpha+\beta$.

A simple algebraic estimate (using both triangle inequality, and $\|\bar\Lambda\|_1=\operatorname{tr}\bar\Lambda=\bar s$) yields the inequality
\begin{equation}
  \|\tilde{\rho}_U-\bar{\rho}\|_1
  \;\le\; \frac{2\big(|\alpha_U-\alpha|+|\beta_U-\beta|\big)+\|\xi_U\|_1}{s_U}.
  \label{eq:master-ineq-final}
\end{equation}

To better understand the contributions to the bound in Eq.~\eqref{eq:master-ineq-final}, we analyzed the statistics of the coefficients $\alpha_U$, $\beta_U$, and the traceless component $\xi_U$ across random Clifford encoders (see Fig.\,\ref{fig:alphabetaXi}).  
The quantities $\alpha_U$ and $\beta_U$ show clear signs of self-averaging: their distributions narrow significantly as the system size increases, and the normalized deviations $(\alpha_U-\alpha)/s_U$ and $(\beta_U-\beta)/s_U$ both decrease rapidly with $N$.  
Numerically, these deviations are consistent with an exponential decay, suggesting that the scalar parts of the decomposition concentrate strongly in the thermodynamic limit.  

In contrast, the behavior of the traceless part $\xi_U$ is more subtle.  
When measured via the trace norm $\|\xi_U\|_1$, the normalized expectation $\mathbb{E}_U[\|\xi_U\|_1/s_U]$ does not exhibit the same rapid decay as the scalar terms, but instead remains roughly constant across the system sizes accessible to our simulations. This can be understood from the fact that $\xi_U$ spans the orthogonal complement to $\{\mathds{1}/d_L, \rho_L\}$, a subspace of dimension $d_L^2-2$, which grows exponentially with the number of logical qubits $k$.  
Fluctuations distributed over such a vast operator space are expected to concentrate only at substantially larger system sizes than we can probe numerically. However, the data show no indication of divergence with $N$; the contribution appears bounded rather than growing, consistent with eventual concentration in the thermodynamic limit.

Taken together, these observations indicate that the leading contributions to the bound, coming from fluctuations of $\alpha_U$ and $\beta_U$, vanish rapidly with system size, while the traceless part is suppressed only much more slowly.  
This explains why at small $N$ the residual term dominates the numerical bound, even though the Hilbert–Schmidt diagnostics in the main text already demonstrate that the full states $\tilde{\rho}_U$ do concentrate around their average $\bar\rho$.  
We therefore conclude that the apparent plateau in the $\xi_U$ contribution reflects finite-size limitations rather than a fundamental obstruction to concentration in the thermodynamic limit.
We emphasize that the concentration is more readily visible in the Hilbert–Schmidt norm than in the trace norm.  
This can be understood from the inequality $\|\rho - \sigma\|_2 \;\le\; \|\rho - \sigma\|_1 \;\le\; \sqrt{d_L}\,\|\rho - \sigma\|_2$, which shows that a small Hilbert–Schmidt distance already implies small trace distance up to a factor $\sqrt{d_L}$.  
Because this factor grows exponentially with the number of logical qubits $k$, detecting decay directly in trace norm requires much larger system sizes than are numerically accessible.  
The Hilbert–Schmidt norm therefore provides a more sensitive diagnostic at moderate $N$: it suppresses fluctuations quickly enough that the onset of concentration can already be seen in our data.  
Importantly, this does not affect the asymptotic conclusion.

\vspace{1cm}

\end{document}